\documentclass[]{aa}  
\usepackage{natbib}
\usepackage{pdflscape}
\usepackage{rotating}
\usepackage{lscape}

\usepackage{amsmath,amssymb }
\usepackage{txfonts}
\usepackage{graphicx, xspace}
\usepackage{multirow}
\usepackage{amssymb}
\usepackage[switch]{lineno}

\newcommand{\num} {$\nu_{\rm max}$\xspace}

\newcommand{\dnu} {$\Delta\nu$\xspace}
\newcommand{\TESS} {TESS\xspace}

\newcommand{\Kepler} {\textit{Kepler}\xspace}

\newcommand{\Hipparcos} {\textit{Hipparcos}\xspace}

\def\dnu{$\Delta\nu$\xspace}
\def\dn1{$\delta\nu_{01}$\xspace}
\def\dn2{$\delta\nu_{02}$\xspace}

\newcommand{\Figure}[1]{Figure\,\ref{#1}\xspace}
\newcommand{\Fig}[1]{Fig.\,\ref{#1}\xspace}

\newcommand{\Table}[1]{Table\,\ref{#1}\xspace}

\usepackage[colorlinks=true,
                        linkcolor=cyan,
                        citecolor=cyan, 
                        filecolor=cyan, 
                        urlcolor=cyan]{hyperref}

\usepackage{color}

\begin{document}

\title{
99 oscillating red-giant stars in binary systems with NASA\,\TESS and NASA\,\textit{Kepler} identified from the SB9-Catalogue}
%\\ with the NASA TESS space telescope}

\titlerunning{Oscillating red-giant stars in binary systems with NASA\,\TESS and NASA\,\Kepler}
\authorrunning{Beck\,et\,al.}

\author{P.\,G.~Beck\inst{\ref{inst:Graz},\ref{inst:IAC}} 
\and S.~Mathur\inst{\ref{inst:IAC},\ref{inst:ULL}}
\and K.~Hambleton\inst{\ref{inst:Philly}}
\and R.\,A.~Garc{\'i}a\inst{\ref{inst:CEA}}
\and L.~Steinwender\inst{\ref{inst:Graz}}
\and \\ N.\,L.~Eisner\inst{\ref{inst:Oxf}} 
\and J.-D.~do~Nascimento,~Jr.\inst{\ref{inst:CfA},\ref{inst:Natal}}
\and P.~Gaulme\inst{\ref{inst:MPS}} 
\and S.~Mathis\inst{\ref{inst:CEA}}
}

\institute{Institut für Physik, Karl-Franzens Universität Graz, Universitätsplatz 5/II, NAWI Graz, 8010 Graz, Austria \label{inst:Graz}.\newline
\email{paul.beck@uni-graz.at}
\and Instituto de Astrof\'{\i}sica de Canarias, E-38200 La Laguna, Tenerife, Spain \label{inst:IAC}
\and Departamento de Astrof\'{\i}sica, Universidad de La Laguna, E-38206 La Laguna, Tenerife, Spain \label{inst:ULL}
\and Department of Astrophysics and Planetary Science, Villanova University, 800 East Lancaster Avenue, Villanova, PA 19085, USA \label{inst:Philly}
\and AIM, CEA, CNRS, 
Universit\'e de Paris
F-91191 Gif-sur-Yvette, France \label{inst:CEA}
\and Department of Physics, University of Oxford, Keble Road, Oxford OX1 3RH, UK \label{inst:Oxf}
\and Center for Astrophysics $|$ Harvard $\&$ Smithsonian, 60 Garden Street, Cambridge, MA 02138, USA \label{inst:CfA} \label{inst:CfA}
\and Universidade Federal do Rio Grande do Norte (UFRN), Departamento de F\'isica, 59078-970, Natal, RN, Brazil \label{inst:Natal}
\and Max-Planck-Institut für Sonnensystemforschung, Justus-von-Liebig-Weg 3, 37077 Göttingen, Germany \label{inst:MPS}
}

\date{Received 24 December 2021 / Accepted 28 March 2022}

\abstract{Oscillating red-giant stars in binary systems are an ideal testbed for investigating the structure and evolution of stars in the advanced phases of evolution. With 83 known red giants in binary systems, of which only $\sim$40 have determined global seismic parameters and orbital parameters, the sample is small compared to the numerous known oscillating stars. The detection of red-giant binary systems is typically obtained from the signature of stellar binarity in space photometry.  
The time base of such data biases the detection towards systems with shorter periods and orbits of insufficient size to allow a red giant to fully extend as it evolves up the red-giant branch. Consequently, the sample shows an excess of H-shell burning giants while containing very few stars in the He-core burning phase.
From the \textit{ninth catalogue of spectroscopic binary orbits} (SB9), we identified candidate systems hosting a red-giant primary component. Searching space photometry from the NASA missions \Kepler, K2, and TESS \textit{(Transiting Exoplanet Survey Satellite)} as well as the BRITE \textit{(BRIght Target Explorer)} constellation mission, we find 99 systems, which were previously unknown to host an oscillating giant component. 
The revised search strategy allowed us to extend the range of orbital periods of systems hosting oscillating giants up to 26\,000\,days. Such wide orbits allow a rich population of He-core burning primaries, which are required for a complete view of stellar evolution from binary studies. Tripling the size of the sample of known oscillating red-giant stars in binary systems is an important step towards an ensemble approach for seismology and tidal studies. While for non-eclipsing binaries the inclination is unknown, such a seismically well-characterized sample will be a treasure trove in combination with \textit{Gaia} astrometric orbits for binary~systems. 
}

\maketitle

%\linenumbers
%\modulolinenumbers[5]

\begin{figure*}[t!]
    \centering
    \includegraphics[width=\textwidth,height=55mm]
    {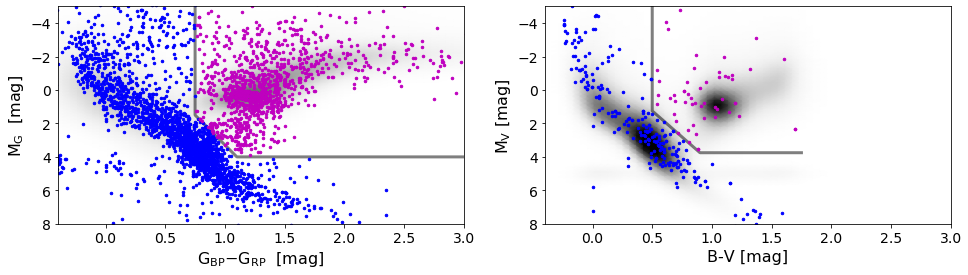}
    \caption{Colour--magnitude diagrams depicting the positions of the systems listed in the SB9 catalogue using \textit{Gaia}\,DR2 (left) and \Hipparcos astrometry (right panel). 
    Magenta dots indicate candidates of hosting primaries in advanced phases of stellar evolution. Blue dots mark the remaining systems. The grey line indicates the selection criterion for red-giant candidates. 
    The background density plots represent the distribution of all stars measured by the respective mission, with a limiting magnitude in V of 10\,mag.}
    \label{fig:photCalibration}
\end{figure*}

\section{Introduction}
Binary systems are gravitationally bound pairs of stars that orbit around a common centre of mass \citep[e.g.][]{Prsa2018}. Unless created in a rare capturing event, both stellar components are born from the same cloud \citep{MoeStefano2017}. Because both stars are located at the same distance and have similar initial conditions and stellar age, we are able to place significant constraints on input parameters for the stellar models. Studying such well-constrained systems offers the opportunity to test complex microscopic and macroscopic physics through \hbox{stellar models \citep[e.g.][]{Johnston2019}.} 

If a stellar component is oscillating, asteroseismology \citep{Asteroseismology1984}, the characterization of the internal structure and dynamics of stars through oscillation modes, can provide crucial, independently inferred  information on the stellar mass, radius, and age of the oscillating components \citep*[see  monograph of][and references therein]{Aerts2010}. 
Binary systems that host solar-like oscillators are of particular interest. Stochastic oscillations driven through convection are found in many objects from solar-like dwarfs to luminous red-giant stars and allow the investigation of a wide range of stellar evolutionary phases with a consistent methodology. 

Despite the large number of more than 16\,000 known solar-like oscillating red-giant (RG) stars \citep{Yu2018, Jackiewicz2021} from the NASA\,\Kepler spacecraft \citep{Borucki2010} alone, there are only 83 known RG oscillators in binary systems. 
This sample was compiled from the analysis of photometric data in a series of papers summarised and referenced in Tables \ref{tab:publishedSampleComplete}, 
\ref{tab:publishedSampleIncomplete}, and  \ref{tab:sampleHardlyAnyInformation}.
%\citet[]{Hekker2010}, 
%\citet[]{Frandsen2013},
%\citet[]{Beck2014a,Beck2015Toulouse,Beck2018Asterix},
%\citet[]{Gaulme2013, Gaulme2014, Gaulme2016, Gaulme2020},
%\citet[]{Rawls2016},
%\citet[]{Brogaard2018}, 
%\citet[]{Themessl2018},
%\cite{GaulmeGuzik2019}, and 
%\citet[]{Benbakoura2021},
Studying these systems led to exciting insights into the evolution of evolved stars, such as the effect of structural changes on the seismic scaling relations \citep{Gaulme2016,Themessl2018,Kallinger2018,Benbakoura2021}, studies of tides \citep{Beck2018Tides}, surface rotation, and activity \citep{Gaulme2014}, calibration of the convective mixing length in low-luminosity giants \citep{LiT2018}, and seismic probing of the first dredge-up event \hbox{and internal rotational gradient \citep{Beck2018Asterix}.}

About 3000 binary stars have been identified in the \Kepler data \citep{Prsa2011,Kirk2016}. The majority of these stars are oriented such that the orbital plane is edge-on and eclipses occur; however, many of these ($\sim$800 in the \Kepler data) present ellipsoidal variations. The majority of ellipsoidal systems have close-to-zero eccentricity due to their small orbits, although 117 of the \Kepler sample are eccentric and colloquially referred to as {heartbeat stars}, a term coined by \cite{Thompson2012} based on the shape of the light curve \citep{kumar1995}. 
However, longer-period non-eclipsing systems of moderate eccentricity are rarely found from time-series photometry. A similar problem exists for eclipsing systems if the orbital periods exceed the  timescale of the time-series. 
Binary systems detected from eclipses observed with space-photometric data are biased towards relatively short orbital periods. For a mission like \Kepler, with a time base of 4\,yr, the majority of the sample extends up to 2-3\,yr.  
Longer periodic systems are normally discovered through radial-velocity (RV) variations from spectroscopic surveys \citep[e.g.][]{Badenes2018}. Coordinated and long-term RV monitoring is then required to determine their orbital parameters. The {ninth catalogue of spectroscopic binary orbits}\footnote{\href{https://sb9.astro.ulb.ac.be/}{https://sb9.astro.ulb.ac.be/}} (SB9) by \cite{Pourbaix2004} provides a compilation of 4004 solved orbits of binary and triple systems across all spectral types and a wide range of orbital periods. 

This paper is structured as follows. In Sect.\,\ref{sec:PhotometricCalibration} we use the SB9 catalogue to search for known binaries that potentially host a RG primary. Section\,\ref{sec:photometry} describes our search of existing space-photometry data of the RG binary candidate system to identify solar-like oscillations. Therefore, we exploit data of the NASA\,\Kepler, its refurbished K2 missions \citep{Howell2014}, and the BRITE (\textit{BRIght Target Explorer}) constellation space telescopes \citep{Weiss2014}. 
For data from the ongoing all-sky survey with \textit{Transiting Exoplanet Survey Satellite}, TESS \citep{Ricker2014} we describe the extraction and postprocessing of the light curves. The seismic analysis of the newly found systems containing an oscillating RG component is explained in Sect.\,\ref{sec:SeismicAnalysis}. In Sects.\,\ref{sec:discussionConlusions} and \ref{sec:desert} we compare the seismic and orbital parameters resulting from this work to the sample in the literature. The work is summarised in Sect.\,\ref{sec:outlook}.

\section{Identifying red giants in SB9 \label{sec:PhotometricCalibration}}
Systems in the SB9 catalogue are typically brighter than 12$^\mathrm{th}$ magnitude. \cite{Pourbaix2004} note that the collection is rich in stars of spectral types later than mid-F, which are the spectral classes where solar-like oscillations are expected. 
First results on RGs from the \TESS mission by \cite{SilvaAguirre2020}, \cite{Mackereth2021}, and \cite{Hon2021} show that this magnitude range is well-suited for detecting RG oscillations with the satellite.

\begin{figure}[t!]
\centering
%\vspace{-5mm}
    \includegraphics[width=\columnwidth,
    height=55mm]{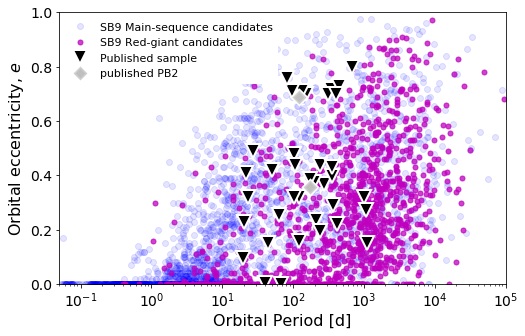}
    \caption{Position of the SB9 binary systems \citep{Pourbaix2004} in the e-P plane. 
    Magenta and blue dots indicate all systems found through the photometric calibration described in Sect.\,\ref{sec:PhotometricCalibration} and   \Fig{fig:photCalibration} to host a primary in an advanced stellar evolutionary phase and on the main sequence, respectively.
    Systems from the published \Kepler sample with an oscillating RG primary are depicted as black triangles. Systems hosting two oscillating RG components (PB2) are shown as grey diamonds.
    }
    \label{fig:epPlane_candidates}
\end{figure}

The search for oscillating RG stars in binary systems started with a photometric calibration of the SB9 catalogue. 
Stars in advanced evolutionary phases were identified through their position on a colour--magnitude diagram (CMD).
For the majority of the systems in the SB9 catalogue, astrometric solutions and multi-colour photometry are available in the \textit{Gaia}\,data\,release\,2 \citep[DR2]{GaiaDR2}. We therefore use the G$_\mathrm{BP}$-G$_\mathrm{RP}$ colour index as a temperature proxy. The provided \textit{Gaia} distance module allows us to calculate the absolute visual magnitude, M$_\mathrm{G}$ (\Fig{fig:photCalibration}, left panel).
For about 230 SB9-systems, no \textit{Gaia} solution exists in DR2, but parallaxes were measured as part of the ESA \Hipparcos mission \citep{vanLeeuwen2007}. For these systems, the classical Johnson B-V colour index was used, and the absolute magnitude was calculated for Johnson V (\Fig{fig:photCalibration}, right panel).

To guide the eye in the respective parameter space in \Fig{fig:photCalibration}, the density distributions of stars brighter than tenth magnitude are shown in the background of the  colour--magnitude diagram (CMD) of the SB9 systems.  Based on this distribution, we set the selection criterion in the parameter space in the CMD to engulf the red-giant branch (RGB), asymptotic-giant branch (AGB), and red clump (RC) and allow for a conservative margin of error in the colour index. In total, we identified 1222 candidate systems potentially hosting a RG. 
We note that such photometric calibration treats the integrated brightness as a single-star source. This simplifying assumption partly explains the larger scatter of the SB9 objects when compared to the well-defined RG phases shown from space data in \Fig{fig:photCalibration}. Furthermore, systems in this regime could also be massive stars, initially OB stars moving horizontally into the super-giant phase. 
The distribution of the 1222 RG candidates in the eccentricity versus orbital-period (e-P) plane is \hbox{illustrated in \Fig{fig:epPlane_candidates}.}

\section{Space photometry \label{sec:photometry}}
\subsection{\TESS}

\longtab{
\begin{landscape}
%\begin{table}
\label{tab:longTable}
\centering

\tabcolsep=2pt
\begin{longtable}{rrr|rrr|rrrrr|rrrcc}%{60mm}

\caption{Catalogue and parameters for oscillating red-giant binaries characterized through photometry from the TESS mission. \label{tab:targetsTESS}}\\

\hline\hline
\multicolumn{1}{c}{SB9} 	&       
\multicolumn{1}{c}{TIC}     &               &       
\multicolumn{1}{c}{Tess}   &       
\multicolumn{1}{c}{V}   	&       
\multicolumn{1}{c}{T$_\mathrm{eff}$}  &       
\multicolumn{1}{c}{$\nu_\mathrm{max}$} &       
\multicolumn{1}{c}{$\Delta\nu$}    	&                       
\multicolumn{1}{c}{M}                   	&       
\multicolumn{1}{c}{R}                   	&       
\multicolumn{1}{c}{$\log$$g$}           	&       
\multicolumn{1}{c}{P$_\mathrm{orbit}$}  &       
\multicolumn{1}{c}{$e$}                 &       
\multicolumn{1}{c}{K$_1$}       &       
\multicolumn{1}{c}{K$_2$}       &       
\multicolumn{1}{c}{SB9} \\

\multicolumn{1}{c}{Seq} &      		 &               &       
\multicolumn{1}{c}{Sect}  &       
\multicolumn{1}{c}{[mag]}       &       
\multicolumn{1}{c}{[K]}                         &       
\multicolumn{1}{c}{[$\mu$Hz]}                   &       
\multicolumn{1}{c}{[$\mu$Hz]}                   &                       
\multicolumn{1}{c}{[M/M$_\odot$]}                       &       
\multicolumn{1}{c}{[R/R$_\odot$]}                       &       
\multicolumn{1}{c}{[dex]}                       &       
\multicolumn{1}{c}{[days]}                      &                               &       
\multicolumn{1}{c}{[km/s]}      &       
\multicolumn{1}{c}{[km/s]}      &       
\multicolumn{1}{c}{Gr}  \\[2mm] \hline
\endfirsthead

%\caption{Catalogue and parameters for oscillating red-giant binaries characterized through photometry from the TESS mission.  (continued) }\\

\hline%\hline
\multicolumn{1}{c}{SB9} 	&       
\multicolumn{1}{c}{TIC}     &               &       
\multicolumn{1}{c}{Tess}   &       
\multicolumn{1}{c}{V}   	&       
\multicolumn{1}{c}{T$_\mathrm{eff}$}  &       
\multicolumn{1}{c}{$\nu_\mathrm{max}$} &       
\multicolumn{1}{c}{$\Delta\nu$}    	&                       
\multicolumn{1}{c}{M}                   	&       
\multicolumn{1}{c}{R}                   	&       
\multicolumn{1}{c}{$\log$$g$}           	&       
\multicolumn{1}{c}{P$_\mathrm{orbit}$}  &       
\multicolumn{1}{c}{$e$}                 &       
\multicolumn{1}{c}{K$_1$}       &       
\multicolumn{1}{c}{K$_2$}       &       
\multicolumn{1}{c}{SB9} \\

\multicolumn{1}{c}{Seq} &      		 &               &       
\multicolumn{1}{c}{Sect}  &       
\multicolumn{1}{c}{[mag]}       &       
\multicolumn{1}{c}{[K]}                         &       
\multicolumn{1}{c}{[$\mu$Hz]}                   &       
\multicolumn{1}{c}{[$\mu$Hz]}                   &                       
\multicolumn{1}{c}{[M/M$_\odot$]}                       &       
\multicolumn{1}{c}{[R/R$_\odot$]}                       &       
\multicolumn{1}{c}{[dex]}                       &       
\multicolumn{1}{c}{[days]}                      &                               &       
\multicolumn{1}{c}{[km/s]}      &       
\multicolumn{1}{c}{[km/s]}      &       
\multicolumn{1}{c}{Gr}  \\[2mm] \hline
\endhead
\hline
\endfoot
%%%%%%%%%%%%%%%%%%%%%%%%%%%%%%%%%%%%%%%%%%%%%%%%%%%

2882    &       49243933                        &               &       3       &       6.87    &       4170    $\pm$   83               &       2.13                    &       0.38                    &                        2.6                     &       66.1                    &       1.21                     &       1509.6  $\pm$   1       &       0.658   $\pm$   0.1     &       6.27    &       $-$     &       $-$     \\
2954    &       243337598                       &               &       9       &       5.91    &       3955    $\pm$   307              &       2.84                    &       0.5                     &                        1.9                     &       50.0                    &       1.32                     &       3856    $\pm$   22      &       0.57    $\pm$   0.01    &       4.1     &       $-$     &       $-$     \\
3869    &       86279245                        &               &       2       &       6.23    &       4173    $\pm$   212              &       3.15                    &       0.6                     &                        1.4                     &       40.1                    &       1.38                     &       184.258 $\pm$   0.023   &       0.102   $\pm$   0.008   &       8.34    &       $-$     &       5       \\
3901    &       233829187                       &               &       2       &       9.68    &       4137    $\pm$   118              &       3.94                    &       0.71                    &                        1.4                     &       35.6                    &       1.47                     &       541     $\pm$   0.8     &       0.043   $\pm$   0.012   &       6.37    &       $-$     &       5       \\
1864    &       302537716                       &               &       2       &       10.10   &       4473    $\pm$   622              &       5.42                    &       0.9                     &                        1.5                     &       31.7                    &       1.63                     &       241.9   $\pm$   1       &       0.04    $\pm$   0.06    &       4.7     &       $-$     &       $-$     \\
2155    &       237415575                       &               &       2       &       7.12    &       4690    $\pm$   287              &       5.59                    &       0.89                    &                        1.9                     &       34.1                    &       1.65                     &       1768    $\pm$   23      &       0                        &       2.58    &       $-$     &       $-$     \\
1869    &       284428523                       &               &       2       &       6.14    &       4256    $\pm$   207              &       5.8                     &       0.89                    &                        1.8                     &       33.6                    &       1.65                     &       2230    $\pm$   23      &       0.14    $\pm$   0.04    &       2.77    &       $-$     &       $-$     \\
958     &       165509120                       &               &       2       &       8.00    &       4223    $\pm$   98               &       6.2                     &       0.98                    &                        1.5                     &       29.6                    &       1.67                     &       794.5                   &       0.69                     &       11.5    &       $-$     &       4       \\
1098    &       236777961                       &               &       13      &       4.64    &       4476    $\pm$   85               &       7.75                    &       1.32                    &                        1.0                     &       21.3                    &       1.78                     &       138.42  $\pm$   0.016   &       0.114   $\pm$   0.014   &       23.46   &       $-$     &       3       \\
1746    &       115581508                       &               &       3       &       6.36    &       4365    $\pm$   138              &       9.11                    &       1.52                    &                        0.9                     &       18.7                    &       1.85                     &       4991    $\pm$   17      &       0.404   $\pm$   0.011   &       4.78    &       $-$     &       5       \\
788     &       383477233                       &               &       1       &       7.59    &       4599    $\pm$   91               &       10.48   $\pm$   0.5     &       1.39    $\pm$   0.59    &                        2.0     $\pm$   1.7     &       25.8    $\pm$   15.6    &       1.92    $\pm$   0.02    &       699.3                    &       0.4                     &       14.9    &       $-$     &       4       \\
2885    &       49253887                        &               &       3       &       7.79    &       4780    $\pm$   64               &       11.65   $\pm$   1.6     &       1.3     $\pm$   1.41    &                        3.7     $\pm$   8.0     &       32.8    $\pm$   50.5    &       1.97    $\pm$   0.06    &       1212.6  $\pm$   2.6     &       0.575   $\pm$   0.007   &       8.18    &       $-$     &       $-$     \\
1568    &       161624708                       &               &       2       &       (J=5.1) &       4910    $\pm$   85               &       11.66   $\pm$   0.82    &       1.64    $\pm$   0.3     &                        1.6     $\pm$   0.6     &       21.6    $\pm$   5.8     &       1.98    $\pm$   0.03    &       6500                     &       0.645   $\pm$   0.028   &       3.49    &       $-$     &       5       \\
2624    &       63445000                        &               &       2       &       7.90    &       4798    $\pm$   54               &       13.07   $\pm$   1.01    &       1.59    $\pm$   0.32    &                        2.4     $\pm$   1.0     &       25.0    $\pm$   7.4     &       2.02    $\pm$   0.03    &       1593    $\pm$   0.6     &       0.6561  $\pm$   0.0021  &       18.11   &       $-$     &       $-$     \\
950     &       85379733                        &               &       2       &       5.07    &       4349    $\pm$   174              &       14.01   $\pm$   0.88    &       2.04    $\pm$   0.21    &                        1.0     $\pm$   0.2     &       15.9    $\pm$   2.6     &       2.03    $\pm$   0.03    &       876.25                   &       0.61                    &       4.8     &       $-$     &       3       \\
1448    &       176753379                       &               &       1       &       7.77    &       4730    $\pm$   98               &       14.09   $\pm$   2.3     &       1.67    $\pm$   0.01    &                        2.4     $\pm$   0.7     &       24.3    $\pm$   3.9     &       2.05    $\pm$   0.07    &       3234                     &       0.71                    &       7.9     &       $-$     &       3       \\
1184    &       87306827                        &               &       1       &       7.87    &       4484    $\pm$   86               &       14.2    $\pm$   0.9     &       1.8     $\pm$   0.16    &                        1.7     $\pm$   0.4     &       20.7    $\pm$   2.9     &       2.05    $\pm$   0.03    &       295.36                   &       0.12                    &       21.7    &       $-$     &       4       \\
229     &       417891254                       &               &       1       &       8.16    &       4324    $\pm$   83               &       14.8    $\pm$   0.66    &       2.17    $\pm$   2.03    &                        0.9     $\pm$   1.7     &       14.8    $\pm$   19.6    &       2.06    $\pm$   0.02    &       488                      &       0.44                    &       4       &       $-$     &       3       \\
1279    &       329613266                       &               &       2       &       7.44    &       4846    $\pm$   434              &       16.57   $\pm$   1.15    &       1.75    $\pm$   0.29    &                        3.3     $\pm$   1.2     &       26.0    $\pm$   6.6     &       2.13    $\pm$   0.04    &       581.4                    &       0.29                    &       8       &       $-$     &       3       \\
3596    &       358152485                       &               &       2       &       6.21    &       4617    $\pm$   92               &       17.14   $\pm$   3.33    &       2.27    $\pm$   0.31    &                        1.3     $\pm$   0.6     &       16.1    $\pm$   4.4     &       2.13    $\pm$   0.08    &       215.3021        $\pm$   0.0665  &       0.124   $\pm$   0.049   &       21.69   &       20.52   &       5       \\
393     &       438006090                       &               &       2       &       7.64    &       4840    $\pm$   315              &       17.68   $\pm$   1.21    &       1.99    $\pm$   0.68    &                        2.5     $\pm$   1.7     &       21.7    $\pm$   10.6    &       2.16    $\pm$   0.03    &       111.69                   &       0.12                    &       7.8     &       $-$     &       3       \\
3920    &       83620534                        &               &       3       &       6.91    &       4623    $\pm$   116              &       18.15   $\pm$   1.11    &       2.08    $\pm$   0.34    &                        2.1     $\pm$   0.7     &       20.0    $\pm$   4.8     &       2.16    $\pm$   0.03    &       1006.9  $\pm$   0.4     &       0.04    $\pm$   0.005   &       8.77    &       $-$     &       5       \\
796     &       357387775                       &               &       1       &       7.89    &       4692    $\pm$   92               &       18.65   $\pm$   1.03    &       2.44    $\pm$   0.42    &                        1.3     $\pm$   0.5     &       15.3    $\pm$   3.8     &       2.17    $\pm$   0.02    &       302.67                   &       0.49                    &       8.6     &       $-$     &       4       \\
1848    &       192775243                       &               &       1       &       7.18    &       4730    $\pm$   102              &       18.97   $\pm$   0.9     &       2.05    $\pm$   0.19    &                        2.6     $\pm$   0.5     &       21.6    $\pm$   3.0     &       2.18    $\pm$   0.02    &       45.779  $\pm$   0.003   &       0                        &       26.91   &       $-$     &       5       \\
1238    &       290519412                       &               &       1       &       6.08    &       4233    $\pm$   106              &       19.2    $\pm$   2.0     &       2.32    $\pm$   0.19    &                        1.4     $\pm$   0.3     &       16.4    $\pm$   2.6     &       2.16    $\pm$   0.04    &       117.776                  &       0.24                    &       22.6    &       $-$     &       4       \\
2331    &       83148763                        &               &       3       &       8.14    &       4648    $\pm$   81               &       19.96   $\pm$   1.03    &       2.46    $\pm$   0.25    &                        1.5     $\pm$   0.3     &       16.0    $\pm$   2.4     &       2.20    $\pm$   0.02    &       9.06    $\pm$   0.0001  &       0                        &       63.79   &       65.98   &       5       \\
1239    &       164660231                       &               &       1       &       5.86    &       4722    $\pm$   59               &       21.34   $\pm$   1.4     &       2.55    $\pm$   0.09    &                        1.6     $\pm$   0.2     &       16.0    $\pm$   1.3     &       2.23    $\pm$   0.03    &       377.6                    &       0.07                    &       11.2    &       $-$     &       2       \\
3486    &       456332026                       &               &       1       &       8.85    &       4833    $\pm$   54               &       21.47   $\pm$   0.94    &       2.9     $\pm$   0.34    &                        1.0     $\pm$   0.3     &       12.7    $\pm$   2.2     &       2.24    $\pm$   0.02    &       13611   $\pm$   109     &       0.41    $\pm$   0.02    &       4       &       $-$     &       $-$     \\
1271    &       468581123                       &               &       1       &       7.04    &       4661    $\pm$   36               &       21.7    $\pm$   2.1     &       2.57    $\pm$   0.24    &                        1.6     $\pm$   0.4     &       15.9    $\pm$   2.6     &       2.24    $\pm$   0.04    &       2871                     &       0.63                    &       3.5     &       $-$     &       4       \\
763     &       27212194                        &               &       1       &       6.44    &       4809    $\pm$   49               &       21.83   $\pm$   1.08    &       2.58    $\pm$   1.68    &                        1.7     $\pm$   2.2     &       16.1    $\pm$   14.8    &       2.25    $\pm$   0.02    &       1366.8                   &       0.19                    &       15.2    &       $-$     &       4       \\
3844    &       362214936                       &               &       2       &       6.39    &       4746    $\pm$   38               &       22.05   $\pm$   1.14    &       2.97    $\pm$   0.36    &                        1.0     $\pm$   0.3     &       12.4    $\pm$   2.2     &       2.25    $\pm$   0.02    &       381.64  $\pm$   0.07    &       0.253   $\pm$   0.005   &       6.67    &       $-$     &       5       \\
403     &       42436617                        &               &       2       &       6.31    &       4903    $\pm$   370              &       22.65   $\pm$   2.38    &       2.23    $\pm$   0.26    &                        3.3     $\pm$   1.0     &       22.1    $\pm$   4.5     &       2.27    $\pm$   0.05    &       428.1                    &       0.26                    &       8       &       $-$     &       3       \\
419     &       234234968                       &               &       2       &       4.47    &       4670    $\pm$   101              &       22.65   $\pm$   2.4     &       2.46    $\pm$   0.48    &                        2.1     $\pm$   0.9     &       17.9    $\pm$   5.3     &       2.26    $\pm$   0.04    &       1760.9                   &       0.4                     &       4       &       $-$     &       3       \\
1576    &       453169877                       &               &       2       &       7.91    &       4745    $\pm$   35               &       22.76   $\pm$   2.82    &       2.5     $\pm$   0.04    &                        2.1     $\pm$   0.4     &       17.6    $\pm$   2.2     &       2.26    $\pm$   0.05    &       1214.257        $\pm$   5.712   &       0.14    $\pm$   0.015   &       4.97    &       $-$     &       5       \\
3924    &       198486253                       &               &       4       &       5.99    &       4816    $\pm$   81               &       24.01   $\pm$   1.54    &       3.04    $\pm$   0.26    &                        1.2     $\pm$   0.2     &       12.9    $\pm$   1.8     &       2.29    $\pm$   0.03    &       2006.4  $\pm$   3.1     &       0.167   $\pm$   0.008   &       4.96    &       $-$     &       5       \\
3983    &       160067845                       &               &       6       &       8.61    &       4649    $\pm$   246              &       29.73   $\pm$   1.5     &       3.41    $\pm$   0.31    &                        1.3     $\pm$   0.3     &       12.4    $\pm$   1.8     &       2.37    $\pm$   0.03    &       26000   $\pm$   5000    &       0.73    $\pm$   0.04    &       5.47    &       5.53    &       5       \\
3904    &       283768251                       &               &       1       &       6.18    &       4975    $\pm$   38               &       29.99   $\pm$   2.62    &       3.5     $\pm$   0.77    &                        1.4     $\pm$   0.6     &       12.3    $\pm$   4.0     &       2.39    $\pm$   0.04    &       878.9   $\pm$   0.6     &       0.21    $\pm$   0.007   &       5.33    &       $-$     &       5       \\
3514    &       159570515                       &               &       4       &       5.96    &       4972    $\pm$   74               &       30.61   $\pm$   3.0     &       3.26    $\pm$   0.56    &                        1.9     $\pm$   0.7     &       14.3    $\pm$   3.8     &       2.40    $\pm$   0.04    &       1778.6  $\pm$   7.8     &       0.501   $\pm$   0.023   &       1.61    &       $-$     &       $-$     \\
1582    &       63798520                        &               &       2       &       9.52    &       4825    $\pm$   92               &       31.6    $\pm$   2.81    &       3.56    $\pm$   0.36    &                        1.4     $\pm$   0.4     &       12.3    $\pm$   2.1     &       2.41    $\pm$   0.04    &       1881.53 $\pm$   18.59   &       0.226   $\pm$   0.036   &       3.08    &       $-$     &       5       \\
2793    &       178871420                       &               &       3       &       8.88    &       4780    $\pm$   67               &       31.75   $\pm$   3.95    &       3.89    $\pm$   0.32    &                        1.0     $\pm$   0.3     &       10.4    $\pm$   1.8     &       2.41    $\pm$   0.05    &       4037    $\pm$   61      &       0.228   $\pm$   0.019   &       10.43   &       $-$     &       $-$     \\
3988    &       207339706                       &               &       4       &       8.73    &       5055    $\pm$   88               &       33.78   $\pm$   3.48    &       3.95    $\pm$   0.3     &                        1.2     $\pm$   0.3     &       11.0    $\pm$   1.6     &       2.45    $\pm$   0.04    &       1324.1  $\pm$   1.1     &       0.103   $\pm$   0.003   &       17.38   &       $-$     &       5       \\
2749    &       143478622                       &               &       3       &       5.57    &       5456    $\pm$   343              &       35.88   $\pm$   0.03    &       3.71    $\pm$   4.07    &                        2.1     $\pm$   4.6     &       13.6    $\pm$   21.1    &       2.49    $\pm$   0.02    &       1212.1  $\pm$   1.1     &       0.715   $\pm$   0.012   &       19.7    &       $-$     &       5       \\\hline \newpage
3461    &       7696886                 &               &       4       &       9.53    &       5131    $\pm$   146              &       36.57   $\pm$   3.5     &       4.54    $\pm$   0.02    &                        0.9     $\pm$   0.2     &       9.2     $\pm$   0.9     &       2.49    $\pm$   0.04    &       5112    $\pm$   63      &       0.84    $\pm$   0.035   &       9.3     &       $-$     &       $-$     \\
2850    &       224598071                       &               &       11      &       6.04    &       4623    $\pm$   138              &       37.1    $\pm$   3.2     &       4.41    $\pm$   0.39    &                        0.9     $\pm$   0.2     &       9.3     $\pm$   1.4     &       2.47    $\pm$   0.04    &       900     $\pm$   14      &       0.07    $\pm$   0.068   &       3.36    &       $-$     &       $-$     \\
1030    &       233078994                       &               &       12      &       7.99    &       4858    $\pm$   70               &       38.6    $\pm$   1.8     &       4.38    $\pm$   0.32    &                        1.2     $\pm$   0.2     &       10.0    $\pm$   1.1     &       2.50    $\pm$   0.02    &       659.4                    &       0.08                    &       6.1     &       $-$     &       3       \\
3775    &       67262475                        &               &       2       &       8.07    &       4945    $\pm$   47               &       38.82   $\pm$   4.09    &       3.98    $\pm$   0.26    &                        1.7     $\pm$   0.4     &       12.2    $\pm$   1.7     &       2.50    $\pm$   0.04    &       5609    $\pm$   55      &       0.22    $\pm$   0.01    &       1.67    &       $-$     &       $-$     \\
302     &       311362722                       &               &       1       &       8.85    &       4500    $\pm$   107              &       39.4    $\pm$   2.9     &       3.38    $\pm$   0.02    &                        2.9     $\pm$   0.4     &       16.0    $\pm$   1.2     &       2.49    $\pm$   0.03    &       3683                     &       0.37                    &       5.2     &       $-$     &       3       \\
3935    &       64366713                        &               &       2       &       6.37    &       4850    $\pm$   74               &       39.39   $\pm$   4.1     &       3.44    $\pm$   0.76    &                        3.0     $\pm$   1.4     &       16.1    $\pm$   5.3     &       2.51    $\pm$   0.04    &       4201    $\pm$   16      &       0                        &       3.68    &       $-$     &       5       \\
1870    &       274060349                       &               &       2       &       8.55    &       4665    $\pm$   70               &       39.54   $\pm$   2.91    &       4.19    $\pm$   0.38    &                        1.4     $\pm$   0.3     &       10.9    $\pm$   1.6     &       2.50    $\pm$   0.03    &       406.9   $\pm$   0.3     &       0.408   $\pm$   0.018   &       10.76   &       $-$     &       $-$     \\
290     &       354428725                       &               &       1       &       8.12    &       4433    $\pm$   156              &       40.29   $\pm$   3.32    &       4.17    $\pm$   0.1     &                        1.4     $\pm$   0.2     &       10.9    $\pm$   1.0     &       2.50    $\pm$   0.04    &       1051                     &       0.04                    &       7       &       $-$     &       3       \\
1569    &       199823594                       &               &       1       &       9.02    &       4501    $\pm$   115              &       40.39   $\pm$   5.15    &       3.74    $\pm$   0.96    &                        2.1     $\pm$   1.2     &       13.6    $\pm$   5.2     &       2.50    $\pm$   0.05    &       1824.286        $\pm$   7.057   &       0.194   $\pm$   0.015   &       4.55    &       $-$     &       5       \\
3959    &       229751007                       &               &       13      &       8.61    &       5017    $\pm$   71               &       42.76   $\pm$   3.37    &       4.9     $\pm$   0.45    &                        1.1     $\pm$   0.2     &       9.1     $\pm$   1.4     &       2.55    $\pm$   0.03    &       435.1   $\pm$   0.4     &       0.075   $\pm$   0.016   &       2.85    &       $-$     &       5       \\
945     &       274676814                       &               &       2       &       6.11    &       4806    $\pm$   118              &       44.96   $\pm$   1.7     &       4.49    $\pm$   0.39    &                        1.6     $\pm$   0.3     &       11.0    $\pm$   1.4     &       2.56    $\pm$   0.02    &       786                      &       0.3                     &       10      &       $-$     &       1       \\
3843    &       298554304                       &               &       5       &       6.30    &       4841    $\pm$   79               &       45      $\pm$   3.46    &       4.75    $\pm$   0.36    &                        1.3     $\pm$   0.3     &       9.9     $\pm$   1.3     &       2.56    $\pm$   0.03    &       1248.5  $\pm$   0.7     &       0.255   $\pm$   0.007   &       5.62    &       $-$     &       5       \\
3823    &       257530105                       &               &       2       &       8.11    &       4883    $\pm$   45               &       45.2    $\pm$   4.9     &       4.88    $\pm$   0.51    &                        1.2     $\pm$   0.3     &       9.5     $\pm$   1.7     &       2.57    $\pm$   0.04    &       1593.2  $\pm$   1.8     &       0.606   $\pm$   0.012   &       2.06    &       $-$     &       5       \\
706     &       357304710                       &               &       1       &       6.74    &       4928    $\pm$   59               &       46.56   $\pm$   3.64    &       4.65    $\pm$   0.11    &                        1.6     $\pm$   0.2     &       10.7    $\pm$   0.9     &       2.58    $\pm$   0.03    &       1355                     &       0.3                     &       5.5     &       $-$     &       4       \\
2841    &       73446226                        &               &       1       &       5.83    &       4668    $\pm$   76               &       48.35   $\pm$   5.7     &       4.97    $\pm$   0.01    &                        1.3     $\pm$   0.3     &       9.5     $\pm$   1.1     &       2.59    $\pm$   0.05    &       629.2   $\pm$   2.7     &       0.135   $\pm$   0.015   &       10.67   &       $-$     &       $-$     \\
1867    &       245663551                       &               &       1       &       7.05    &       4717    $\pm$   229              &       49.47   $\pm$   4.09    &       4.52    $\pm$   0.21    &                        2.0     $\pm$   0.4     &       11.7    $\pm$   1.3     &       2.60    $\pm$   0.04    &       3061    $\pm$   19      &       0.22    $\pm$   0.03    &       4.75    &       $-$     &       $-$     \\
3892    &       318232318                       &               &       1       &       6.73    &       4830    $\pm$   81               &       49.57   $\pm$   2.24    &       4.97    $\pm$   0.93    &                        1.4     $\pm$   0.6     &       9.9     $\pm$   2.7     &       2.60    $\pm$   0.02    &       300.33  $\pm$   0.06    &       0                        &       5.76    &       $-$     &       5       \\
1572    &       250141410                       &               &       2       &       7.99    &       4979    $\pm$   79               &       49.69   $\pm$   5.71    &       5.16    $\pm$   0.37    &                        1.3     $\pm$   0.3     &       9.4     $\pm$   1.4     &       2.61    $\pm$   0.05    &       1263.23 $\pm$   3.729   &       0.395   $\pm$   0.025   &       6.24    &       $-$     &       5       \\
2653    &       139124331                       &               &       2       &       8.46    &       4880    $\pm$   122              &       50.71   $\pm$   4.65    &       4.15    $\pm$   0.21    &                        3.1     $\pm$   0.6     &       14.3    $\pm$   1.7     &       2.62    $\pm$   0.04    &       1136.6  $\pm$   0.5     &       0.226   $\pm$   0.007   &       10.33   &       $-$     &       $-$     \\
2825    &       467473356                       &               &       2       &       7.14    &       4753    $\pm$   108              &       56.31   $\pm$   3.38    &       6.07    $\pm$   0.5     &                        0.9     $\pm$   0.2     &       7.6     $\pm$   1.0     &       2.66    $\pm$   0.03    &       1815.3  $\pm$   1.2     &       0.503   $\pm$   0.0024  &       11.37   &       12.31   &       $-$     \\
3921    &       428980427                       &               &       1       &       5.76    &       4885    $\pm$   269              &       57.01   $\pm$   5.65    &       5.05    $\pm$   0.21    &                        2.0     $\pm$   0.4     &       11.0    $\pm$   1.3     &       2.67    $\pm$   0.04    &       1034    $\pm$   1.1     &       0.403   $\pm$   0.011   &       2.9     &       $-$     &       5       \\
1895    &       250394854                       &               &       1       &       5.94    &       5053    $\pm$   142              &       58.0    $\pm$   5.7     &       4.74    $\pm$   0.45    &                        2.8     $\pm$   0.7     &       12.8    $\pm$   2.1     &       2.68    $\pm$   0.04    &       2066.1  $\pm$   2.6     &       0.366   $\pm$   0.005   &       12.41   &       $-$     &       $-$     \\
1880    &       293286598                       &               &       1       &       6.30    &       5131    $\pm$   57               &       58.02   $\pm$   4.83    &       5.07    $\pm$   0.34    &                        2.3     $\pm$   0.4     &       11.4    $\pm$   1.4     &       2.69    $\pm$   0.03    &       5302    $\pm$   18      &       0.432   $\pm$   0.013   &       8.71    &       $-$     &       $-$     \\
3020    &       71595072                        &               &       1       &       9.16    &       4835    $\pm$   90               &       60.93   $\pm$   4.11    &       5.66    $\pm$   0.23    &                        1.6     $\pm$   0.2     &       9.4     $\pm$   0.8     &       2.69    $\pm$   0.03    &       469.74  $\pm$   0.07    &       0.602   $\pm$   0.005   &       9.53    &       $-$     &       4       \\
3612    &       289966090                       &               &       2       &       5.92    &       4998    $\pm$   54               &       61.4    $\pm$   5.1     &       4.97    $\pm$   0.24    &                        2.8     $\pm$   0.5     &       12.2    $\pm$   1.3     &       2.71    $\pm$   0.03    &       2681.08 $\pm$   0.62    &       0.735   $\pm$   0.0024  &       8.66    &       $-$     &       5       \\
3996    &       198161444                       &               &       14      &       9.24    &       4984    $\pm$   193              &       64.7    $\pm$   3.2     &       5.64    $\pm$   0.40    &                        2.0     $\pm$   0.3     &       10.1    $\pm$   1.2     &       2.73    $\pm$   0.02    &       3027    $\pm$   5       &       0.457   $\pm$   0.008   &       4.87    &       $-$     &       5       \\
3104    &       284596799                       &               &       1       &       6.49    &       5103    $\pm$   61               &       66.16   $\pm$   4.09    &       6.29    $\pm$   0.49    &                        1.5     $\pm$   0.3     &       8.5     $\pm$   1.1     &       2.74    $\pm$   0.03    &       923.3   $\pm$   1.1     &       0                        &       4.23    &       $-$     &       $-$     \\
3460    &       52942882                        &               &       2       &       7.30    &       5221    $\pm$   119              &       78.15   $\pm$   4.58    &       7.9     $\pm$   0.07    &                        1.0     $\pm$   0.1     &       6.5     $\pm$   0.4     &       2.82    $\pm$   0.03    &       1641    $\pm$   24      &       0.352   $\pm$   0.018   &       9.05    &       $-$     &       $-$     \\
3837    &       367865950                       &               &       4       &       5.97    &       5160    $\pm$   112              &       79.5    $\pm$   3.8     &       7.20    $\pm$   0.51    &                        1.5     $\pm$   0.2     &       7.8     $\pm$   0.9     &       2.82    $\pm$   0.02    &       2962.3  $\pm$   0.5     &       0.831   $\pm$   0.001   &       14.37   &       $-$     &       5       \\
3989    &       198213983                       &               &       14      &       8.72    &       5145    $\pm$   127              &       83.6    $\pm$   4.1     &       7.35    $\pm$   0.57    &                        1.6     $\pm$   0.3     &       7.9     $\pm$   1.0     &       2.85    $\pm$   0.02    &       1016.6  $\pm$   0.9     &       0.185   $\pm$   0.01    &       4.42    &       $-$     &       5       \\
1840    &       359634548                       &               &       12      &       7.43    &       4966    $\pm$   59               &       84.8    $\pm$   5.8     &       7.17    $\pm$   0.51    &                        1.7     $\pm$   0.3     &       8.2     $\pm$   1.0     &       2.84    $\pm$   0.03    &       4456    $\pm$   17      &       0.488   $\pm$   0.016   &       5.05    &       $-$     &       4       \\
3103    &       53250605                        &               &       1       &       3.72    &       4710    $\pm$   82               &       86.18   $\pm$   7.92    &       6.87    $\pm$   0.06    &                        1.9     $\pm$   0.3     &       8.8     $\pm$   0.8     &       2.84    $\pm$   0.04    &       1283.4  $\pm$   0.7     &       0.231   $\pm$   0.017   &       2.28    &       $-$     &       $-$     \\
210     &       391085379                       &               &       1       &       7.16    &       4740    $\pm$   89               &       87.64   $\pm$   6.65    &       8.04    $\pm$   0.6     &                        1.1     $\pm$   0.2     &       6.7     $\pm$   0.9     &       2.85    $\pm$   0.03    &       927                      &       0.38                    &       3.4     &       $-$     &       4       \\
2843    &       284858336                       &               &       2       &       6.68    &       4918    $\pm$   58               &       119.14  $\pm$   5.6     &       9.8     $\pm$   0.61    &                        1.4     $\pm$   0.2     &       6.2     $\pm$   0.6     &       2.99    $\pm$   0.02    &       930     $\pm$   11      &       0.114   $\pm$   0.05    &       8.31    &       $-$     &       $-$     \\[2mm] \hline
                                                                                                                                                                                                         
3839    &       \multirow{2}{*}{275691882}                      &       o       &       \multirow{2}{*}{14}     &       \multirow{2}{*}{8.24}   &       \multirow{2}{*}{5106$\pm$65}                            &       \multirow{2}{*}{66.15$\pm$3.74}                 &       \multirow{2}{*}{6.49$\pm$0.06}                  &                       \multirow{2}{*}{1.3$\pm$0.1}                    &       \multirow{2}{*}{8$\pm$0.5}                      &       \multirow{2}{*}{2.74$\pm$0.02}                  &       1293    $\pm$   1.2     &       0.4798  $\pm$   0.0031  &       14.7    &       15.16   &       5       \\
3840    &                               &       i       &               &               &                                       &                               &                               &                                               &                               &                               &       688.4   $\pm$   0.9     &       0.476   $\pm$   0.013   &       3.85    &       $-$     &       5       \\ \hline
\end{longtable}
\end{landscape}
}

The existing full-frame image (FFI) data of the \TESS mission up to Sector 40 were searched for all objects on the candidate list using the {\textsc{Astropy}} package \citep{astropy2018}. Photometric time-series for each RG candidate were extracted from the FFI data using the {\textsc{Eleanor}}-package \citep{Feinstein2019}. 

\textsc{Eleanor} optimises the extracted light curve for the detection and analysis of exoplanet transit signals. Transit detection requires the highest achievable signal-to-noise ratio (S/N) possible, and therefore a small aperture around the best-illuminated pixels. \cite{Garcia2014} demonstrated that asteroseismic investigation requires robust light curves, which are achieved by larger apertures. Consequently, we extracted the data, forcing a larger aperture than the optimal aperture, determined by \textsc{Eleanor}. \TESS FFI data had a cadence of 30 minutes. This was increased to 10\,min in the extended mission (years 3-4). To reduce the scatter in the data, we rebinned the 10\,min cadence \hbox{to the classical 30\,min cadence.}

We selected 77 systems (\Table{tab:targetsTESS}) for which the visual inspection of the power spectrum density (PSD) showed the presence of the acoustic-mode bump below the Nyquist frequency of 283\,$\mu$Hz corresponding to the 30-min cadence for detailed seismic analysis (Sect.\,\ref{sec:SeismicAnalysis}).
To improve the spectral window of the data, gaps of up to two days in the light curve were filled through an inpainting technique \citep{Garcia2014, Pires2015}. 
Unfortunately, none of the systems in which solar-like oscillations were found were eclipsing. This might be due to the long orbital periods compared to the relatively short \hbox{times-series provided by \TESS.} 

\begin{figure}
    \centering
    \includegraphics[width=\columnwidth]
    {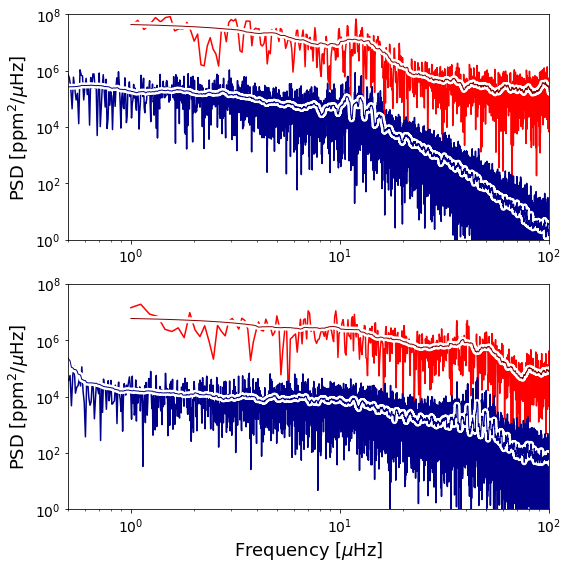}
    \caption{Power-spectral density for two oscillating RG primaries in binary systems. The PSDs of \Kepler and TESS data are shown in blue and red, respectively. The smoothed power is depicted as a solid line. For better comparison, the TESS-PSD is shifted by two orders \hbox{of magnitude.}}
    \label{fig:TargetsInCommon}
\end{figure}

\begin{figure*}[t!]
\centering
    \includegraphics[width=0.85\textwidth]%,height=75mm]
    {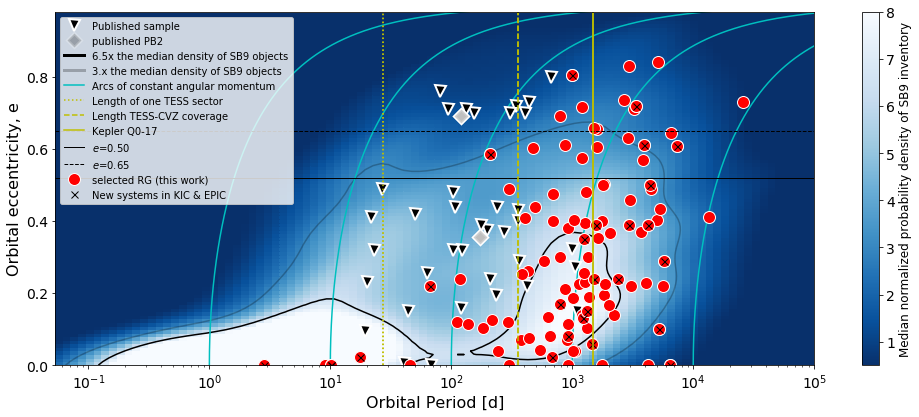}
    \caption{Distribution of the sample of binary systems hosting evolved stars as their primary stellar component. 
    Red dots mark systems seismically characterised in this study. 
    Systems from the published \Kepler sample with an oscillating RG primary are depicted as black triangles. Systems hosting two oscillating RG components (PB2) are shown as grey diamonds.
    The background density plot represents the distribution of all 4004 systems for which the SB9 \citep{Pourbaix2004} reports orbital solutions. The grey and black contour lines engulf the regions with 3 and 6.5 times the median density of a bin.
    The light-blue lines indicate the arcs of constant angular momentum in the e-P plane for circular orbital periods for 1, 10, 100, 1000, and 10000 days. 
    The vertical yellow lines indicate the time-span covered by \TESS and \Kepler. The horizontal black lines indicate the range of eccentricities that are not generally found in systems with periods shorter than 200\,d.
    }
    \label{fig:PeriodEccPlane}
\end{figure*}

\subsection{\Kepler and K2}
Similarly, we cross-correlated the candidate list with the NASA\,\Kepler data archives, and the re-purposed K2 mission data. This search revealed 16 and 3 known binary systems, respectively, with oscillating RG primaries (\Table{tab:targetsKepler}). 
These light curves were treated as identical to the \TESS data. Calibrated light curves and their corresponding PSD are available on the MAST archive\footnote{\href{https://archive.stsci.edu/prepds/kepseismic/}{https://archive.stsci.edu/prepds/kepseismic/}}.

\TESS visited the \Kepler field for 1 or 2 sectors during the second year of its operation. Figure\,\ref{fig:TargetsInCommon} compares two successful detections by both satellites. As expected, the resolution of four years of \Kepler photometry is superior to a few months of \TESS observations. Because the K2 fields are located close to the ecliptic, \TESS has not yet observed these targets.

\subsection{BRITE}
Using BRITE data, \cite{Kallinger2019} were able to seismically characterise 23 RG stars, either through a direct detection of the oscillation-power excess or by measuring the granulation timescales. Among them are two systems that are listed in SB9, \object{39\,Cyg}, and \object{12\,Mus}.

\section{Asteroseismic analysis \label{sec:SeismicAnalysis}}

While the TESS sample was selected from visual inspection of the PSD, this was not the case for the Kepler and K2 targets. Among the Kepler and K2 sample, 14 targets and 3 targets respectively exhibit solar-like oscillations.
The oscillation signature in the PSD of a solar-like oscillator, as depicted in \Fig{fig:TargetsInCommon}, is sufficiently characterised  through the global seismic parameters \citep[e.g.][and references therein]{Aerts2010}. The central frequency of the excess oscillation power, \num , is determined from a Gaussian envelope, which is simultaneously fit with two power laws and a constant offset to describe the granulation signal, the photon noise, and the background, respectively \citep[e.g.][]{Carrier2010}. The large-frequency separation, \dnu, between modes of the same spherical degree but consecutive radial order \citep{Tassoul1980} is determined from the power spectrum of the PSD in the frequency range of the excess of the oscillation power \citep[e.g.][]{Mathur2010}. We analysed the \textsc{Eleanor} light curves, processed with the standard corrections. For targets with non-significant mode detections, we analysed the light curves, corrected through the principal component analysis (PCA) to remove common instrumental systematic errors.
For all targets, we determined the global seismic parameters using the A2Z pipeline \citep[][]{Mathur2010}.

Using the asteroseismic scaling relations \citep[]{Kjeldsen1995, Chaplin2011}, the measured \num and \dnu, complemented by the measured effective temperature, T$_\mathrm{eff}$ , allow us to infer the mass and radius for the oscillating star. If available, the effective temperatures were taken from APOGEE \citep{Apogee2020Ahumada}. For the remaining targets, the effective temperature was taken from the \textit{Gaia}\,DR2 catalogue, whereby the quoted uncertainty was estimated from the given upper and lower temperature range of the solution. 
The correction between the asymptotic and observed large-frequency separation was obtained following \cite{Mosser2013}.

The global seismic parameters and the obtained masses and radii for the \TESS, \Kepler, and K2 sample are presented in \Table{tab:targetsTESS}, \ref{tab:targetsKepler}, respectively. For stars with \num$\leq$10\,$\mu$Hz, we only provide an indicative value for \num and \dnu. Indeed, in this frequency range the detected modes are not in the asymptotic regime (i.e. low n/l), and therefore the measured large-frequency separation cannot be directly used in seismic scaling relations.
The low frequency resolution resulting from the relatively short timescale covered by TESS observations further complicates the determination of the densely packed power excess at such low frequencies. This increases the uncertainties on the inferred seismic quantities.

\Table{tab:BriteRv} reports the seismic values from the literature for the targets of BRITE and RV studies. Table\,\ref{tab:targetsK2nonosc} presents an additional 15 RG candidate systems, which were identified in our photometric search and with existing \Kepler or K2 data. However, no oscillations were found in these systems. Work by \cite{Gaulme2014} showed that non-oscillating systems typically have strong spot-modulated light curves, providing valuable input for the tidal modelling \citep{Beck2018Tides}. We therefore report these systems for the completeness of the sample.

\section{Distribution of seismic and orbital parameters \label{sec:discussionConlusions}}

The seismic Hertzsprung-Russell diagram (HRD) in Fig.\,\ref{fig:seismicHRD} compares the newly established sample of oscillating RG stars in binary systems to the 82 systems of the literature sample (see Appendix\,\ref{appendix:publishedSample}).  Table\,\ref{tab:publishedSampleComplete} presents 39 previously known systems with determined global seismic parameters and known orbital periods and eccentricities. 
Table\,\ref{tab:publishedSampleIncomplete} presents 20 previously known systems for which only the global seismic parameters for the RG primary are known. 
Table\,\ref{tab:sampleHardlyAnyInformation} lists 23 previously known systems with their very limited known parameters. By adding 99 additional systems to the literature, we have more than doubled the size of the known sample and more than tripled the sample of stars with known global seismic parameters and orbital period and eccentricity. 
%\vspace{-3mm}

\subsection{Seismic characterisation of the sample}

\begin{figure}[t!]
\centering
%\vspace{-5mm}
    \includegraphics[width=\columnwidth]{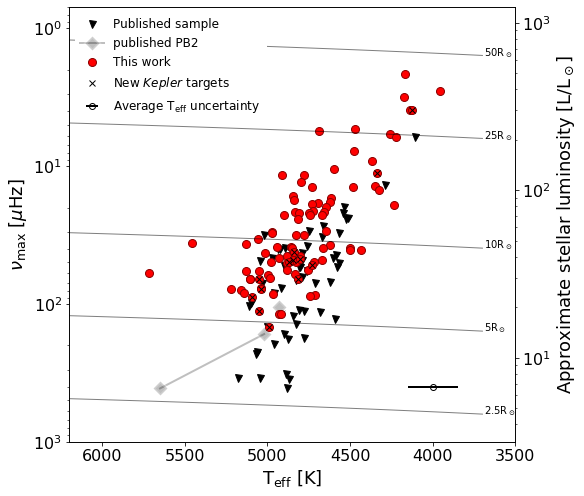}
    \caption{Distribution of the sample in the seismic HRD with the frequency of the oscillation-power excess on the left vertical axis. The symbols are identical to \Fig{fig:PeriodEccPlane}. 
    The bottom marker indicates the average uncertainty of the effective temperature of 150\,K.
    \label{fig:seismicHRD}}
\end{figure}

The distribution of the systems in  Fig.\,\ref{fig:seismicHRD} shows that the literature sample mainly populated the region of 30\,$\mu$Hz\,$\leq$\,\num$\,\leq$\,400\,$\mu$Hz. The new sample ($\sim$2\,$\mu$Hz\,$\leq$\,\num$\leq$\,100\,$\mu$Hz) presented in this work extends the sample of oscillating RG binaries to lower oscillation frequencies and therefore to the more luminous regime of stars similar to luminous giants like Aldebaran \citep[$\sim$2\,$\mu$Hz,][]{Beck2020}.
While the analysis of the mixed-mode patterns is beyond the scope of this paper, the position of the new sample in the seismic HRD (Fig.\,\ref{fig:seismicHRD}) and the wide orbital periods (\Fig{fig:PeriodEccPlane}) suggest that the sample is also rich in He-core burning stars, which have successfully undergone ignition of their helium core at the tip of the RGB.  This lifts the bias on the evolutionary status found in the previous samples.

It is worth pointing out that eight systems are of the same age, as they are confirmed members of the open cluster \object{NGC\,6819}, which were discovered in a spectroscopic survey by \cite{Milliman2014}. Their membership as well as the cluster age of $\sim$2.3\,Gyr were obtained through asteroseismology by \cite{Stello2011Clusters}, \cite{Basu2011}, \hbox{and \cite{Handberg2017}.}

\subsection{Distribution of the orbital parameters}
The distribution of the orbital parameters of the systems in the e--P plane is depicted in \Fig{fig:PeriodEccPlane}. 
The background contour plot illustrates the distribution of all orbits listed in the SB9 catalogue. The majority of systems have periods below $\sim$100 days with low-eccentricities or even circularised orbits ($e$\,$\lesssim$\,0.2). This overdensity is mostly populated with hot stars below the spectral type F, whose structure is dominated by radiative \hbox{regions \citep{Torres2010AccurateMasses}.}

All candidate RG systems, which were identified from the CMD in Sect.\,\ref{sec:PhotometricCalibration} are shown as small magenta dots in \Fig{fig:PeriodEccPlane}. Uncertainties are not available for the first systems listed in the SB9 catalogue, but are provided for more recent entries.
While a few RG systems share the same region in the parameter space as the bulk of the hot binaries, most of the evolved stars are found in systems with orbital periods 200\,d\,$\lesssim$\,P$_\mathrm{orb}$\,$\lesssim$\,3000\,d, and $e$\,$\lesssim$\,0.6. This rich sample of evolved stars form a second, tear-drop-shaped overdensity  found in the distribution of the SB9 systems. In addition, in this period regime, the SB9 catalogue contains systems that cover the entire possible range, from circularised orbits ($e$\,=\,0) to a reported maximum of $e_\mathrm{max}$\,=\,0.972. 

\Figure{fig:PeriodEccPlane} illustrates the differences in the orbital parameters between the sample collected from the literature and systems analyzed in this work. Both samples show a similar distribution in the covered range of eccentricities.  
Stars in the \Kepler sample were selected because of their photometric binary signatures, such as periodically occurring eclipses or heartbeat events. Such selection constraints favour systems with periods shorter than the duration of the mission (shown as yellow vertical lines in \Fig{fig:PeriodEccPlane}). Because a RG will expand its radius by two orders of magnitude as it advances on the RGB, it is unlikely that systems with such periods will survive the unavoidable common-envelope phase unaltered. 
Therefore, it is not surprising that the \Kepler sample predominantly contains stars in the rapidly evolving H-shell-burning RGB phase \citep{Beck2018Tides} and contains a lower number of stars quiescently burning He in the core and maintaining a stable luminosity. From the full catalogue of published systems in App.\,\ref{appendix:publishedSample}, we find that 38 RGB and 13 RC systems were identified.

Selecting systems based on a priori knowledge of their spectroscopically inferred binary nature removes the bias on the period that was previously introduced by the observation length of space photometry from our sample of binary systems characterised through asteroseismology. This selection also removes the bias introduced by the selection criterion of eclipsing systems in space photometry, which also depends on the inclination angle. 
Therefore, systems with very long periods can be studied by asteroseismology with the single requirement that the photometric time-series allows for a frequency resolution sufficient to perform asteroseismic analysis. Only for very close systems, namely of typically P\,$\lesssim$\,20 days, will the magnetic activity ---which is increased because of the acceleration of rotation triggered by tides that strengthen the dynamo action through a tidally driven dynamo--- suppress solar-like oscillations \citep{Gaulme2016, Mathur2019, Benbakoura2021}. Extending the sample with systems with periods of several hundred to thousands of days opens the analysis to wider binary systems whose primary likely has already progressed through the RGB and is now quiescently burning helium in its core.

\section{A red-giant desert in the $e$-P plane? \label{sec:desert}}

\begin{table*}[t!]
    \centering
    \caption{Periastron distance and radius of the Roche lobe between both components of a hypothetical system.}
    \tabcolsep=15pt

    \begin{tabular}{c|cc|cc|cc}
        \hline\hline
         $e$ & \multicolumn{2}{c}{P$_\mathrm{orbit}$=150\,d} 
         & \multicolumn{2}{c}{P$_\mathrm{orbit}$=500\,d}  & \multicolumn{2}{c}{P$_\mathrm{orbit}$=1000\,d} \\
         ~[\,]~& S$_\mathrm{peri}$ [R$_\odot$]& R$_\mathrm{L,1}^\mathrm{Peri}$ [R$_\odot$]  & S$_\mathrm{peri}$ [R$_\odot$]& R$_\mathrm{L,1}^\mathrm{Peri}$ [R$_\odot$]& S$_\mathrm{peri}$ [R$_\odot$]& R$_\mathrm{L,1}^\mathrm{Peri}$ [R$_\odot$]\\ 
        \hline
          0.2   & 123.6 & 50.8  & 275.8 & 113.4  & 437.9 & 180.0   \\ 
          0.4   &  92.7 & 38.1  & 206.9 & 85.0  & 328.4 & 135.0   \\ 
          0.6   &  61.8 & 25.4  & 137.9 & 56.7  & 218.9 & 90.0   \\ 
          0.8   &  30.9 & 12.7  &  68.9 & 28.3  & 109.3 & 44.9   \\ 
        \hline
    \end{tabular}
    \tablefoot{A hypothetical binary system of M$_1$\,=\,1.3\,M$_\odot$ and M$_2$\,=\,0.9\,M$_\odot$ is assumed.  The table reports the theoretical minimum separation between both components S$_\mathrm{peri}$ and the corresponding radius of the Roche lobe for the primary (R$_\mathrm{L,1}^\mathrm{Peri}$) component for a grid in period (P$_\mathrm{orb}:$ 150, 500, and 1000\,d) and orbital eccentricity ($e$: 0.2, 0.4, 0.6, and 0.8) during periastron. All values are given in solar units.}
    \label{tab:perioastronDistances}
\end{table*}

It was pointed out by \cite{Beck2018Tides} and \cite{Benbakoura2021} that the \Kepler sample of RG binaries  does not contain systems with $0.5$\,$\lesssim$\,$e$\,$\lesssim$\,0.7 (for P$_\mathrm{orb}\lesssim200$\,d). 
This apparent gap, which we refer to as the {RG desert}, poses an interesting problem.

The small number of RG binaries in the \Kepler sample (black triangles in Figs.\,\ref{fig:epPlane_candidates} and \ref{fig:PeriodEccPlane}) hindered further conclusions.  The \Kepler sample (black triangles) only extends to orbital periods of about 400\,d, which makes the feature appear more pronounced but is affected by small number statistics. This study increases the number of stars identified in advanced evolutionary stages by two orders of magnitude, which extends to much larger eccentricities. However, the region P$_\mathrm{orb}$\,$\lesssim$\,200\,d, and $e$\,$\gtrsim$\,0.5 remains sparsely populated, while this range of P$_\mathrm{orb}$\,$\gtrsim$\,200\,d is now well populated with RG binary systems.

Tidal circularisation reflects the interplay of two bodies and the dissipation of the kinetic energy of tidal flows into heat. Unless a third body is acting on the system or mass is lost, tidal forces should reduce eccentricity. Tidal theory therefore does not predict such a gap. Spiraling of the secondary onto the primary of the system would occur when the total angular momentum of the orbital movement is less than three times the total angular momentum from the stellar rotation \citep{Hut1980}, which is not the case for these~systems.

The long-periodic edge of the gap at $\sim$200\,d is explained by the fact that the timescales of the tidally driven circularisation are becoming too long-lived to show immediate effects \citep{Beck2018Tides}.
This region in the e-P diagram (P$_\mathrm{orb}$\,$\lesssim$\,200\,d, $e$\,$\gtrsim$\,0.5) is also less populated by stars on the main sequence, as can be seen from the iso-contour of the total distribution.
The bottom edge at $e$\,$\simeq$0.5 coincides with the sudden drop in the population in the e--P plane. 

Detection bias may also affect the distribution. One possible bias is the detectability of a binary system from spectroscopy. With an increasing eccentricity and longer orbital period, the duration in which the binary system shows negligible variations of its radial velocities is growing. Unless measured to a high RV precision, such systems could go unnoticed from a short-term sampling survey. This is particularly relevant if we are facing the long side of the ellipse. Yet, such an argument would instead explain a lower occurrence rate at longer periods. However, the RG desert is found at periods compatible with the typical length of an observing season. Also, even a relatively sparse sampling could reveal a binary system.

Contrary to what might be expected from the large number of RG systems in the SB9, there are about as many systems found from \Kepler as listed in the SB9. This might still point to a selection bias as those systems were selected due to their heartbeat signature in the light curve.
Human bias might explain this. Systems in the RG desert are typically heartbeat stars. While one has to make the effort of selecting them for RV monitoring, their photometric signature simply is a byproduct of light-curve analysis of space data.

Table\,\ref{tab:perioastronDistances} describes the separation and the radius of the Roche lobe \citep[in the formulation of][]{Eggleton1983} around the giant primary between two components of a hypothetical binary system consisting of a primary and secondary of 1.3 and 0.9\,M$_\odot$ on a 150-, 500-, and 1000-day orbit with four different eccentricities of 0.2, 0.4, 0.6, and 0.8. 
Provided that a star at the tip of the RGB reaches a maximum radius of $\sim$170\,R$_\odot$, it can be seen that at some point, the giant will fill its Roche lobe. The mass transfer onto the secondary during this phase of stable Roche lobe overflow \citep{han2002} will drive a rapid decrease in the orbital eccentricity. Supposing that the modified system with lower eccentricity is still too small for the maximum radius of the giant primary, the system will undergo a common-envelope phase, which most likely leads to the ejection of the system and the creation of a hot subdwarf\,B star \citep{han2002}.

All published systems from the \Kepler sample found above (i.e. $e$\,$\gtrsim$\,0.4, P$_\mathrm{orb}$\,$\lesssim$\,200\,d) the RG desert were indeed identified as H-shell burning stars \citep{Beck2018Tides}. This supports the idea that the region was deserted of RG stars due to the expanding stellar radius.
Finally, we note that at this stage this is a purely phenomenological description of the distribution of RG binaries in the $e$--P plane. Further interpretation would require a statistical analysis, which is beyond the current focus of this paper.

\section{Outlook \label{sec:outlook}} 
With this work, we substantially increase the number of known RG stars in binary systems characterised using space photometry. Because of the way in which we selected the sample, that is, only based on SB9 and CMD, most of our systems do not present eclipses in the light curves. 
The unconstrained inclination of the orbit along the line of sight therefore poses limitations for further exploiting the full potential of the ensemble. 
Because the scaling relations treat the oscillating component as a single source, the astroseismic investigation of the RG components delivers inclination-independent masses and radii.  Specifically, through constraints set by binarity on the initial conditions and through accurate orbital solutions, detailed modelling of the complex internal physics will be possible \citep{White2013, Beck2018Asterix}. 

For studies of tidal forces, which scale with the sixth power of the radius \citep{Zahn2013}, such large ensembles of systems with seismically inferred radii and clear identification of their evolutionary state in a wide range of eccentricities will provide important test cases. 
As tidal interactions strengthen the dynamo, we also expect to find many spotted stars, which will allow us to study synchronisation and angular momentum transport in stars.

When the astrometric time-series from \textit{Gaia} arrive, we predict that we will be able to infer the inclination from the shape of the projected binary orbit and the known orbital eccentricity from spectroscopic solutions for wider binaries. 
This growing catalogue of seismically characterised systems is therefore an important step in preparing ensemble asteroseismology RG binaries. 
The approach presented in this paper also allows a targeted search for oscillators in binary systems with the forthcoming ESA \textsc{Plato} mission \citep{Rauer2014}. The multicolour photometry of this mission will be of high value to the study of oscillations and binary modelling.

Extending the sample coverage to longer periods and enriching the sample with He-core burning stars will therefore  vastly improve our ability to study open questions of stellar structure and its evolution.

\vspace{2mm}

\begin{acknowledgements}
We thank the referee for useful comments that allowed us to improve the article.  
PGB supports the financial support by NAWI Graz and acknowledges the support by the \textit{Dr.\,Heinrich-Jörg\,Foundation} at the faculty of natural sciences at Graz University. 
SM acknowledges support by the Spanish Ministry of Science and Innovation with the \textit{Ramon\,y\,Cajal} fellowship number RYC-2015-17697 and the grant number PID2019-107187GB-I00. RAG and StM acknowledge support from the PLATO CNES grant. This paper includes data collected with the \emph{Kepler}\,\&\,\TESS missions, obtained from the MAST data archive at the Space Telescope Science Institute (STScI). KH acknowledges support through NASA ADAP grants (80NSSC19K0594). Funding for these missions is provided by the NASA Science Mission Directorate and by the NASA Explorer Program respectively. STScI is operated by the Association of Universities for Research in Astronomy, Inc., under NASA contract NAS 5–26555. Based on data collected by the BRITE Constellation satellite mission, designed, built, launched, operated and supported by the Austrian Research Promotion Agency (FFG), the University of Vienna, the Technical University of Graz, the University of Innsbruck, the Canadian Space Agency (CSA), the University of Toronto Institute for Aerospace Studies (UTIAS), the Foundation for Polish Science \& Technology (FNiTP MNiSW), and National Science Centre (NCN).
\\
\textit{Software:} \texttt{Python} \citep{10.5555/1593511}, 
\texttt{numpy} \citep{numpy,Harris_2020},  
\texttt{matplotlib} \citep{4160265},  
\texttt{scipy} \citep{2020SciPy-NMeth}, 
\texttt{astropy }\citep{astropy:2013, astropy:2018}.
%\texttt{pandas} \citep{reback2020pandas, mckinney-proc-scipy-2010},
This research made use of Astropy, a community-developed core Python package for Astronomy.

\end{acknowledgements}

\bibliographystyle{aa}
\bibliography{AA202143005.bib}

\begin{thebibliography}{72}
\expandafter\ifx\csname natexlab\endcsname\relax\def\natexlab#1{#1}\fi

\bibitem[{{Aerts} {et~al.}(2010){Aerts}, {Christensen-Dalsgaard}, \&
  {Kurtz}}]{Aerts2010}
{Aerts}, C., {Christensen-Dalsgaard}, J., \& {Kurtz}, D.~W. 2010,
  {Asteroseismology}

\bibitem[{{Ahumada} {et~al.}(2020){Ahumada}, {Prieto}, {Almeida}, {Anders},
  {Anderson}, {Andrews}, {Anguiano}, {Arcodia}, {Armengaud}, {Aubert}, {Avila},
  {Avila-Reese}, {Badenes}, {Balland}, {Barger}, {Barrera-Ballesteros}, {Basu},
  {Bautista}, {Beaton}, {Beers}, {Benavides}, {Bender}, {Bernardi}, {Bershady},
  {Beutler}, {Bidin}, {Bird}, {Bizyaev}, {Blanc}, {Blanton}, {Boquien},
  {Borissova}, {Bovy}, {Brandt}, {Brinkmann}, {Brownstein}, {Bundy}, {Bureau},
  {Burgasser}, {Burtin}, {Cano-D{\'\i}az}, {Capasso}, {Cappellari}, {Carrera},
  {Chabanier}, {Chaplin}, {Chapman}, {Cherinka}, {Chiappini}, {Doohyun Choi},
  {Chojnowski}, {Chung}, {Clerc}, {Coffey}, {Comerford}, {Comparat}, {da
  Costa}, {Cousinou}, {Covey}, {Crane}, {Cunha}, {Ilha}, {Dai}, {Damsted},
  {Darling}, {Davidson}, {Davies}, {Dawson}, {De}, {de la Macorra}, {De Lee},
  {Queiroz}, {Deconto Machado}, {de la Torre}, {Dell'Agli}, {du Mas des
  Bourboux}, {Diamond-Stanic}, {Dillon}, {Donor}, {Drory}, {Duckworth},
  {Dwelly}, {Ebelke}, {Eftekharzadeh}, {Davis Eigenbrot}, {Elsworth},
  {Eracleous}, {Erfanianfar}, {Escoffier}, {Fan}, {Farr},
  {Fern{\'a}ndez-Trincado}, {Feuillet}, {Finoguenov}, {Fofie},
  {Fraser-McKelvie}, {Frinchaboy}, {Fromenteau}, {Fu}, {Galbany}, {Garcia},
  {Garc{\'\i}a-Hern{\'a}ndez}, {Oehmichen}, {Ge}, {Maia}, {Geisler}, {Gelfand},
  {Goddy}, {Gonzalez-Perez}, {Grabowski}, {Green}, {Grier}, {Guo}, {Guy},
  {Harding}, {Hasselquist}, {Hawken}, {Hayes}, {Hearty}, {Hekker}, {Hogg},
  {Holtzman}, {Horta}, {Hou}, {Hsieh}, {Huber}, {Hunt}, {Chitham}, {Imig},
  {Jaber}, {Angel}, {Johnson}, {Jones}, {J{\"o}nsson}, {Jullo}, {Kim},
  {Kinemuchi}, {Kirkpatrick}, {Kite}, {Klaene}, {Kneib}, {Kollmeier}, {Kong},
  {Kounkel}, {Krishnarao}, {Lacerna}, {Lan}, {Lane}, {Law}, {Le Goff}, {Leung},
  {Lewis}, {Li}, {Lian}, {Lin}, {Long}, {Longa-Pe{\~n}a}, {Lundgren}, {Lyke},
  {Ted Mackereth}, {MacLeod}, {Majewski}, {Manchado}, {Maraston}, {Martini},
  {Masseron}, {Masters}, {Mathur}, {McDermid}, {Merloni}, {Merrifield},
  {M{\'e}sz{\'a}ros}, {Miglio}, {Minniti}, {Minsley}, {Miyaji}, {Mohammad},
  {Mosser}, {Mueller}, {Muna}, {Mu{\~n}oz-Guti{\'e}rrez}, {Myers}, {Nadathur},
  {Nair}, {Nandra}, {do Nascimento}, {Nevin}, {Newman}, {Nidever}, {Nitschelm},
  {Noterdaeme}, {O'Connell}, {Olmstead}, {Oravetz}, {Oravetz}, {Osorio},
  {Pace}, {Padilla}, {Palanque-Delabrouille}, {Palicio}, {Pan}, {Pan},
  {Parker}, {Paviot}, {Peirani}, {Ram{\'r}ez}, {Penny}, {Percival},
  {Perez-Fournon}, {P{\'e}rez-R{\`a}fols}, {Petitjean}, {Pieri},
  {Pinsonneault}, {Poovelil}, {Povick}, {Prakash}, {Price-Whelan}, {Raddick},
  {Raichoor}, {Ray}, {Rembold}, {Rezaie}, {Riffel}, {Riffel}, {Rix}, {Robin},
  {Roman-Lopes}, {Rom{\'a}n-Z{\'u}{\~n}iga}, {Rose}, {Ross}, {Rossi},
  {Rowlands}, {Rubin}, {Salvato}, {S{\'a}nchez}, {S{\'a}nchez-Menguiano},
  {S{\'a}nchez-Gallego}, {Sayres}, {Schaefer}, {Schiavon}, {Schimoia},
  {Schlafly}, {Schlegel}, {Schneider}, {Schultheis}, {Schwope}, {Seo},
  {Serenelli}, {Shafieloo}, {Shamsi}, {Shao}, {Shen}, {Shetrone}, {Shirley},
  {Aguirre}, {Simon}, {Skrutskie}, {Slosar}, {Smethurst}, {Sobeck}, {Sodi},
  {Souto}, {Stark}, {Stassun}, {Steinmetz}, {Stello}, {Stermer},
  {Storchi-Bergmann}, {Streblyanska}, {Stringfellow}, {Stutz}, {Su{\'a}rez},
  {Sun}, {Taghizadeh-Popp}, {Talbot}, {Tayar}, {Thakar}, {Theriault}, {Thomas},
  {Thomas}, {Tinker}, {Tojeiro}, {Toledo}, {Tremonti}, {Troup}, {Tuttle},
  {Unda-Sanzana}, {Valentini}, {Vargas-Gonz{\'a}lez}, {Vargas-Maga{\~n}a},
  {V{\'a}zquez-Mata}, {Vivek}, {Wake}, {Wang}, {Weaver}, {Weijmans}, {Wild},
  {Wilson}, {Wilson}, {Wolthuis}, {Wood-Vasey}, {Yan}, {Yang}, {Y{\`e}che},
  {Zamora}, {Zarrouk}, {Zasowski}, {Zhang}, {Zhao}, {Zhao}, {Zheng}, {Zheng},
  {Zhu}, \& {Zou}}]{Apogee2020Ahumada}
{Ahumada}, R., {Prieto}, C.~A., {Almeida}, A., {et~al.} 2020, \apjs, 249, 3

\bibitem[{{Astropy Collab.} {et~al.}(2018){Astropy Collab.}, {Price-Whelan},
  {Sip{\H{o}}cz}, {G{\"u}nther}, {Lim}, {Crawford}, {Conseil}, {Shupe},
  {Craig}, {Dencheva}, {Ginsburg}, {VanderPlas}, {Bradley},
  {P{\'e}rez-Su{\'a}rez}, {de Val-Borro}, {Aldcroft}, {Cruz}, {Robitaille},
  {Tollerud}, {Ardelean}, {Babej}, {Bach}, {Bachetti}, {Bakanov}, {Bamford},
  {Barentsen}, {Barmby}, {Baumbach}, {Berry}, {Biscani}, {Boquien}, {Bostroem},
  {Bouma}, {Brammer}, {Bray}, {Breytenbach}, {Buddelmeijer}, {Burke},
  {Calderone}, {Cano Rodr{\'\i}guez}, {Cara}, {Cardoso}, {Cheedella}, {Copin},
  {Corrales}, {Crichton}, {D'Avella}, {Deil}, {Depagne}, {Dietrich}, {Donath},
  {Droettboom}, {Earl}, {Erben}, {Fabbro}, {Ferreira}, {Finethy}, {Fox},
  {Garrison}, {Gibbons}, {Goldstein}, {Gommers}, {Greco}, {Greenfield},
  {Groener}, {Grollier}, {Hagen}, {Hirst}, {Homeier}, {Horton}, {Hosseinzadeh},
  {Hu}, {Hunkeler}, {Ivezi{\'c}}, {Jain}, {Jenness}, {Kanarek}, {Kendrew},
  {Kern}, {Kerzendorf}, {Khvalko}, {King}, {Kirkby}, {Kulkarni}, {Kumar},
  {Lee}, {Lenz}, {Littlefair}, {Ma}, {Macleod}, {Mastropietro}, {McCully},
  {Montagnac}, {Morris}, {Mueller}, {Mumford}, {Muna}, {Murphy}, {Nelson},
  {Nguyen}, {Ninan}, {N{\"o}the}, {Ogaz}, {Oh}, {Parejko}, {Parley}, {Pascual},
  {Patil}, {Patil}, {Plunkett}, {Prochaska}, {Rastogi}, {Reddy Janga},
  {Sabater}, {Sakurikar}, {Seifert}, {Sherbert}, {Sherwood-Taylor}, {Shih},
  {Sick}, {Silbiger}, {Singanamalla}, {Singer}, {Sladen}, {Sooley},
  {Sornarajah}, {Streicher}, {Teuben}, {Thomas}, {Tremblay}, {Turner},
  {Terr{\'o}n}, {van Kerkwijk}, {de la Vega}, {Watkins}, {Weaver}, {Whitmore},
  {Woillez}, {Zabalza}, \& {Astropy Contributors}}]{astropy2018}
{Astropy Collab.}, {Price-Whelan}, A.~M., {Sip{\H{o}}cz}, B.~M., {et~al.} 2018,
  \aj, 156, 123

\bibitem[{{Astropy Collaboration} {et~al.}(2018){Astropy Collaboration},
  {Price-Whelan}, {Sip{H{o}}cz}, {G{"u}nther}, {Lim}, {Crawford}, {Conseil},
  {Shupe}, {Craig}, {Dencheva}, {Ginsburg}, {Vand erPlas}, {Bradley},
  {P{'e}rez-Su{'a}rez}, {de Val-Borro}, {Aldcroft}, {Cruz}, {Robitaille},
  {Tollerud}, {Ardelean}, {Babej}, {Bach}, {Bachetti}, {Bakanov}, {Bamford},
  {Barentsen}, {Barmby}, {Baumbach}, {Berry}, {Biscani}, {Boquien}, {Bostroem},
  {Bouma}, {Brammer}, {Bray}, {Breytenbach}, {Buddelmeijer}, {Burke},
  {Calderone}, {Cano Rodr{'i}guez}, {Cara}, {Cardoso}, {Cheedella}, {Copin},
  {Corrales}, {Crichton}, {D'Avella}, {Deil}, {Depagne}, {Dietrich}, {Donath},
  {Droettboom}, {Earl}, {Erben}, {Fabbro}, {Ferreira}, {Finethy}, {Fox},
  {Garrison}, {Gibbons}, {Goldstein}, {Gommers}, {Greco}, {Greenfield},
  {Groener}, {Grollier}, {Hagen}, {Hirst}, {Homeier}, {Horton}, {Hosseinzadeh},
  {Hu}, {Hunkeler}, {Ivezi{'c}}, {Jain}, {Jenness}, {Kanarek}, {Kendrew},
  {Kern}, {Kerzendorf}, {Khvalko}, {King}, {Kirkby}, {Kulkarni}, {Kumar},
  {Lee}, {Lenz}, {Littlefair}, {Ma}, {Macleod}, {Mastropietro}, {McCully},
  {Montagnac}, {Morris}, {Mueller}, {Mumford}, {Muna}, {Murphy}, {Nelson},
  {Nguyen}, {Ninan}, {N{"o}the}, {Ogaz}, {Oh}, {Parejko}, {Parley}, {Pascual},
  {Patil}, {Patil}, {Plunkett}, {Prochaska}, {Rastogi}, {Reddy Janga},
  {Sabater}, {Sakurikar}, {Seifert}, {Sherbert}, {Sherwood-Taylor}, {Shih},
  {Sick}, {Silbiger}, {Singanamalla}, {Singer}, {Sladen}, {Sooley},
  {Sornarajah}, {Streicher}, {Teuben}, {Thomas}, {Tremblay}, {Turner},
  {Terr{'o}n}, {van Kerkwijk}, {de la Vega}, {Watkins}, {Weaver}, {Whitmore},
  {Woillez}, {Zabalza}, \& {Astropy Contributors}}]{astropy:2018}
{Astropy Collaboration}, {Price-Whelan}, A.~M., {Sip{H{o}}cz}, B.~M., {et~al.}
  2018, aj, 156, 123

\bibitem[{{Astropy Collaboration} {et~al.}(2013){Astropy Collaboration},
  {Robitaille}, {Tollerud}, {Greenfield}, {Droettboom}, {Bray}, {Aldcroft},
  {Davis}, {Ginsburg}, {Price-Whelan}, {Kerzendorf}, {Conley}, {Crighton},
  {Barbary}, {Muna}, {Ferguson}, {Grollier}, {Parikh}, {Nair}, {Unther},
  {Deil}, {Woillez}, {Conseil}, {Kramer}, {Turner}, {Singer}, {Fox}, {Weaver},
  {Zabalza}, {Edwards}, {Azalee Bostroem}, {Burke}, {Casey}, {Crawford},
  {Dencheva}, {Ely}, {Jenness}, {Labrie}, {Lim}, {Pierfederici}, {Pontzen},
  {Ptak}, {Refsdal}, {Servillat}, \& {Streicher}}]{astropy:2013}
{Astropy Collaboration}, {Robitaille}, T.~P., {Tollerud}, E.~J., {et~al.} 2013,
  \aap, 558, A33

\bibitem[{{Badenes} {et~al.}(2018){Badenes}, {Mazzola}, {Thompson}, {Covey},
  {Freeman}, {Walker}, {Moe}, {Troup}, {Nidever}, {Allende Prieto}, {Andrews},
  {Barb{\'a}}, {Beers}, {Bovy}, {Carlberg}, {De Lee}, {Johnson}, {Lewis},
  {Majewski}, {Pinsonneault}, {Sobeck}, {Stassun}, {Stringfellow}, \&
  {Zasowski}}]{Badenes2018}
{Badenes}, C., {Mazzola}, C., {Thompson}, T.~A., {et~al.} 2018, \apj, 854, 147

\bibitem[{{Basu} {et~al.}(2011){Basu}, {Grundahl}, {Stello}, {Kallinger},
  {Hekker}, {Mosser}, {Garc{\'\i}a}, {Mathur}, {Brogaard}, {Bruntt}, {Chaplin},
  {Gai}, {Elsworth}, {Esch}, {Ballot}, {Bedding}, {Gruberbauer}, {Huber},
  {Miglio}, {Yildiz}, {Kjeldsen}, {Christensen-Dalsgaard}, {Gilliland},
  {Fanelli}, {Ibrahim}, \& {Smith}}]{Basu2011}
{Basu}, S., {Grundahl}, F., {Stello}, D., {et~al.} 2011, \apjl, 729, L10

\bibitem[{{Beck} {et~al.}(2014){Beck}, {Hambleton}, {Vos}, {Kallinger},
  {Bloemen}, {Tkachenko}, {Garc{\'{\i}}a}, {{\O}stensen}, {Aerts}, {Kurtz}, {De
  Ridder}, {Hekker}, {Pavlovski}, {Mathur}, {De Smedt}, {Derekas}, {Corsaro},
  {Mosser}, {Van Winckel}, {Huber}, {Degroote}, {Davies}, {Pr{\v s}a},
  {Debosscher}, {Elsworth}, {Nemeth}, {Siess}, {Schmid}, {P{\'a}pics}, {de
  Vries}, {van Marle}, {Marcos-Arenal}, \& {Lobel}}]{Beck2014a}
{Beck}, P.~G., {Hambleton}, K., {Vos}, J., {et~al.} 2014, A\&A, 564, A36

\bibitem[{{Beck} {et~al.}(2015{\natexlab{a}}){Beck}, {Hambleton}, {Vos},
  {Kallinger}, {Garcia}, {Mathur}, \& {Houmani}}]{Beck2015Toulouse}
{Beck}, P.~G., {Hambleton}, K., {Vos}, J., {et~al.} 2015{\natexlab{a}}, EpjConf
  101, 06004

\bibitem[{{Beck} {et~al.}(2018{\natexlab{a}}){Beck}, {Kallinger}, {Pavlovski},
  {Palacios}, {Tkachenko}, {Mathis}, {Garc{\'{\i}}a}, {Corsaro}, {Johnston},
  {Mosser}, {Ceillier}, {do Nascimento}, \& {Raskin}}]{Beck2018Asterix}
{Beck}, P.~G., {Kallinger}, T., {Pavlovski}, K., {et~al.} 2018{\natexlab{a}},
  \aap, 612, A22

\bibitem[{{Beck} {et~al.}(2015{\natexlab{b}}){Beck}, {Kambe}, {Hillen},
  {Corsaro}, {Van Winckel}, {Moravveji}, {De Ridder}, {Bloemen}, {Saesen},
  {Mathias}, {Degroote}, {Kallinger}, {Verhoelst}, {Ando}, {Carrier}, {Acke},
  {Oreiro}, {Miglio}, {Eggenberger}, {Sato}, {Zwintz}, {P{\'a}pics},
  {Marcos-Arenal}, {Sans Fuentes}, {Schmid}, {Waelkens}, {{\O}stensen},
  {Matthews}, {Yoshida}, {Izumiura}, {Koyano}, {Nagayama}, {Shimizu}, {Okada},
  {Okita}, {Sakamoto}, {Yamamuro}, \& {Aerts}}]{Beck2015a}
{Beck}, P.~G., {Kambe}, E., {Hillen}, M., {et~al.} 2015{\natexlab{b}}, A\&A,
  573, A138

\bibitem[{{Beck} {et~al.}(2020){Beck}, {Kuschnig}, {Houdek}, {Kallinger},
  {Weiss}, {Palle}, {Grundahl}, {Hatzes}, {Parviainen}, {Allende Prieto},
  {Deeg}, {Jim{\'e}nez}, {Mathur}, {Garcia}, {White}, {Bedding}, {Grossmann},
  {Janisch}, {Zaqarashvili}, {Hanslmeier}, \& {Zwintz}}]{Beck2020}
{Beck}, P.~G., {Kuschnig}, R., {Houdek}, G., {et~al.} 2020, in Stars and their
  Variability Observed from Space, ed. C.~{Neiner}, W.~W. {Weiss}, D.~{Baade},
  R.~E. {Griffin}, C.~C. {Lovekin}, \& A.~F.~J. {Moffat}, 75--79

\bibitem[{{Beck} {et~al.}(2018{\natexlab{b}}){Beck}, {Mathis}, {Gallet},
  {Charbonnel}, {Benbakoura}, {Garc{\'\i}a}, \& {do
  Nascimento}}]{Beck2018Tides}
{Beck}, P.~G., {Mathis}, S., {Gallet}, F., {et~al.} 2018{\natexlab{b}}, \mnras,
  479, L123

\bibitem[{{Benbakoura} {et~al.}(2021){Benbakoura}, {Gaulme}, {McKeever},
  {Sekaran}, {Beck}, {Spada}, {Jackiewicz}, {Mathis}, {Mathur}, {Tkachenko}, \&
  {Garc{\'\i}a}}]{Benbakoura2021}
{Benbakoura}, M., {Gaulme}, P., {McKeever}, J., {et~al.} 2021, \aap, 648, A113

\bibitem[{{Borucki} {et~al.}(2010){Borucki}, {Koch}, {Basri}, {Batalha},
  {Brown}, {Caldwell}, {Caldwell}, {Christensen-Dalsgaard}, {Cochran},
  {DeVore}, {Dunham}, {Dupree}, {Gautier}, {Geary}, {Gilliland}, {Gould},
  {Howell}, {Jenkins}, {Kondo}, {Latham}, {Marcy}, {Meibom}, {Kjeldsen},
  {Lissauer}, {Monet}, {Morrison}, {Sasselov}, {Tarter}, {Boss}, {Brownlee},
  {Owen}, {Buzasi}, {Charbonneau}, {Doyle}, {Fortney}, {Ford}, {Holman},
  {Seager}, {Steffen}, {Welsh}, {Rowe}, {Anderson}, {Buchhave}, {Ciardi},
  {Walkowicz}, {Sherry}, {Horch}, {Isaacson}, {Everett}, {Fischer}, {Torres},
  {Johnson}, {Endl}, {MacQueen}, {Bryson}, {Dotson}, {Haas}, {Kolodziejczak},
  {Van Cleve}, {Chandrasekaran}, {Twicken}, {Quintana}, {Clarke}, {Allen},
  {Li}, {Wu}, {Tenenbaum}, {Verner}, {Bruhweiler}, {Barnes}, \&
  {Prsa}}]{Borucki2010}
{Borucki}, W.~J., {Koch}, D., {Basri}, G., {et~al.} 2010, Science, 327, 977

\bibitem[{{Brogaard} {et~al.}(2018){Brogaard}, {Hansen}, {Miglio}, {Slumstrup},
  {Frandsen}, {Jessen-Hansen}, {Lund}, {Bossini}, {Thygesen}, {Davies},
  {Chaplin}, {Arentoft}, {Bruntt}, {Grundahl}, \& {Handberg}}]{Brogaard2018}
{Brogaard}, K., {Hansen}, C.~J., {Miglio}, A., {et~al.} 2018, \mnras, 476, 3729

\bibitem[{{Carrier} {et~al.}(2010){Carrier}, {De Ridder}, {Baudin}, {Barban},
  {Hatzes}, {Hekker}, {Kallinger}, {Miglio}, {Montalb{\'a}n}, {Morel}, {Weiss},
  {Auvergne}, {Baglin}, {Catala}, {Michel}, \& {Samadi}}]{Carrier2010}
{Carrier}, F., {De Ridder}, J., {Baudin}, F., {et~al.} 2010, A\&A, 509, A73

\bibitem[{{Chaplin} {et~al.}(2011){Chaplin}, {Kjeldsen},
  {Christensen-Dalsgaard}, {Basu}, {Miglio}, {Appourchaux}, {Bedding},
  {Elsworth}, {Garc{\'{\i}}a}, {Gilliland}, {Girardi}, {Houdek}, {Karoff},
  {Kawaler}, {Metcalfe}, {Molenda-{\.Z}akowicz}, {Monteiro}, {Thompson},
  {Verner}, {Ballot}, {Bonanno}, {Brand{\~a}o}, {Broomhall}, {Bruntt},
  {Campante}, {Corsaro}, {Creevey}, {Do{\u g}an}, {Esch}, {Gai}, {Gaulme},
  {Hale}, {Handberg}, {Hekker}, {Huber}, {Jim{\'e}nez}, {Mathur}, {Mazumdar},
  {Mosser}, {New}, {Pinsonneault}, {Pricopi}, {Quirion}, {R{\'e}gulo},
  {Salabert}, {Serenelli}, {Silva Aguirre}, {Sousa}, {Stello}, {Stevens},
  {Suran}, {Uytterhoeven}, {White}, {Borucki}, {Brown}, {Jenkins}, {Kinemuchi},
  {Van Cleve}, \& {Klaus}}]{Chaplin2011}
{Chaplin}, W.~J., {Kjeldsen}, H., {Christensen-Dalsgaard}, J., {et~al.} 2011,
  Science, 332, 213

\bibitem[{{Christensen-Dalsgaard}(1984)}]{Asteroseismology1984}
{Christensen-Dalsgaard}, J. 1984, in Space Research in Stellar Activity and
  Variability, ed. {A.~Mangeney \& F.~Praderie}, 11

\bibitem[{{Eggleton}(1983)}]{Eggleton1983}
{Eggleton}, P.~P. 1983, \apj, 268, 368

\bibitem[{{Feinstein} {et~al.}(2019){Feinstein}, {Montet}, {Foreman-Mackey},
  {Bedell}, {Saunders}, {Bean}, {Christiansen}, {Hedges}, {Luger}, {Scolnic},
  \& {Cardoso}}]{Feinstein2019}
{Feinstein}, A.~D., {Montet}, B.~T., {Foreman-Mackey}, D., {et~al.} 2019,
  \pasp, 131, 094502

\bibitem[{{Frandsen} {et~al.}(2013){Frandsen}, {Lehmann}, {Hekker},
  {Southworth}, {Debosscher}, {Beck}, {Hartmann}, {Pigulski}, {Kopacki},
  {Ko{\l}aczkowski}, {St{\c e}{\'s}licki}, {Thygesen}, {Brogaard}, \&
  {Elsworth}}]{Frandsen2013}
{Frandsen}, S., {Lehmann}, H., {Hekker}, S., {et~al.} 2013, A\&A, 556, A138

\bibitem[{{Gaia Collaboration} {et~al.}(2018){Gaia Collaboration}, {Brown},
  {Vallenari}, {Prusti}, {de Bruijne}, {Babusiaux}, {Bailer-Jones}, {Biermann},
  {Evans}, {Eyer}, {Jansen}, {Jordi}, {Klioner}, {Lammers}, {Lindegren},
  {Luri}, {Mignard}, {Panem}, {Pourbaix}, {Randich}, {Sartoretti}, {Siddiqui},
  {Soubiran}, {van Leeuwen}, {Walton}, {Arenou}, {Bastian}, {Cropper},
  {Drimmel}, {Katz}, {Lattanzi}, {Bakker}, {Cacciari}, {Casta{\~n}eda},
  {Chaoul}, {Cheek}, {De Angeli}, {Fabricius}, {Guerra}, {Holl}, {Masana},
  {Messineo}, {Mowlavi}, {Nienartowicz}, {Panuzzo}, {Portell}, {Riello},
  {Seabroke}, {Tanga}, {Th{\'e}venin}, {Gracia-Abril}, {Comoretto},
  {Garcia-Reinaldos}, {Teyssier}, {Altmann}, {Andrae}, {Audard},
  {Bellas-Velidis}, {Benson}, {Berthier}, {Blomme}, {Burgess}, {Busso},
  {Carry}, {Cellino}, {Clementini}, {Clotet}, {Creevey}, {Davidson}, {De
  Ridder}, {Delchambre}, {Dell'Oro}, {Ducourant},
  {Fern{\'a}ndez-Hern{\'a}ndez}, {Fouesneau}, {Fr{\'e}mat}, {Galluccio},
  {Garc{\'\i}a-Torres}, {Gonz{\'a}lez-N{\'u}{\~n}ez}, {Gonz{\'a}lez-Vidal},
  {Gosset}, {Guy}, {Halbwachs}, {Hambly}, {Harrison}, {Hern{\'a}ndez},
  {Hestroffer}, {Hodgkin}, {Hutton}, {Jasniewicz}, {Jean-Antoine-Piccolo},
  {Jordan}, {Korn}, {Krone-Martins}, {Lanzafame}, {Lebzelter}, {L{\"o}ffler},
  {Manteiga}, {Marrese}, {Mart{\'\i}n-Fleitas}, {Moitinho}, {Mora}, {Muinonen},
  {Osinde}, {Pancino}, {Pauwels}, {Petit}, {Recio-Blanco}, {Richards},
  {Rimoldini}, {Robin}, {Sarro}, {Siopis}, {Smith}, {Sozzetti}, {S{\"u}veges},
  {Torra}, {van Reeven}, {Abbas}, {Abreu Aramburu}, {Accart}, {Aerts},
  {Altavilla}, {{\'A}lvarez}, {Alvarez}, {Alves}, {Anderson}, {Andrei},
  {Anglada Varela}, {Antiche}, {Antoja}, {Arcay}, {Astraatmadja}, {Bach},
  {Baker}, {Balaguer-N{\'u}{\~n}ez}, {Balm}, {Barache}, {Barata}, {Barbato},
  {Barblan}, {Barklem}, {Barrado}, {Barros}, {Barstow}, {Bartholom{\'e}
  Mu{\~n}oz}, {Bassilana}, {Becciani}, {Bellazzini}, {Berihuete}, {Bertone},
  {Bianchi}, {Bienaym{\'e}}, {Blanco-Cuaresma}, {Boch}, {Boeche}, {Bombrun},
  {Borrachero}, {Bossini}, {Bouquillon}, {Bourda}, {Bragaglia}, {Bramante},
  {Breddels}, {Bressan}, {Brouillet}, {Br{\"u}semeister}, {Brugaletta},
  {Bucciarelli}, {Burlacu}, {Busonero}, {Butkevich}, {Buzzi}, {Caffau},
  {Cancelliere}, {Cannizzaro}, {Cantat-Gaudin}, {Carballo}, {Carlucci},
  {Carrasco}, {Casamiquela}, {Castellani}, {Castro-Ginard}, {Charlot},
  {Chemin}, {Chiavassa}, {Cocozza}, {Costigan}, {Cowell}, {Crifo}, {Crosta},
  {Crowley}, {Cuypers}, {Dafonte}, {Damerdji}, {Dapergolas}, {David}, {David},
  {de Laverny}, {De Luise}, {De March}, {de Martino}, {de Souza}, {de Torres},
  {Debosscher}, {del Pozo}, {Delbo}, {Delgado}, {Delgado}, {Di Matteo},
  {Diakite}, {Diener}, {Distefano}, {Dolding}, {Drazinos}, {Dur{\'a}n},
  {Edvardsson}, {Enke}, {Eriksson}, {Esquej}, {Eynard Bontemps}, {Fabre},
  {Fabrizio}, {Faigler}, {Falc{\~a}o}, {Farr{\`a}s Casas}, {Federici},
  {Fedorets}, {Fernique}, {Figueras}, {Filippi}, {Findeisen}, {Fonti},
  {Fraile}, {Fraser}, {Fr{\'e}zouls}, {Gai}, {Galleti}, {Garabato},
  {Garc{\'\i}a-Sedano}, {Garofalo}, {Garralda}, {Gavel}, {Gavras}, {Gerssen},
  {Geyer}, {Giacobbe}, {Gilmore}, {Girona}, {Giuffrida}, {Glass}, {Gomes},
  {Granvik}, {Gueguen}, {Guerrier}, {Guiraud}, {Guti{\'e}rrez-S{\'a}nchez},
  {Haigron}, {Hatzidimitriou}, {Hauser}, {Haywood}, {Heiter}, {Helmi}, {Heu},
  {Hilger}, {Hobbs}, {Hofmann}, {Holland}, {Huckle}, {Hypki}, {Icardi},
  {Jan{\ss}en}, {Jevardat de Fombelle}, {Jonker}, {Juh{\'a}sz}, {Julbe},
  {Karampelas}, {Kewley}, {Klar}, {Kochoska}, {Kohley}, {Kolenberg},
  {Kontizas}, {Kontizas}, {Koposov}, {Kordopatis}, {Kostrzewa-Rutkowska},
  {Koubsky}, {Lambert}, {Lanza}, {Lasne}, {Lavigne}, {Le Fustec}, {Le
  Poncin-Lafitte}, {Lebreton}, {Leccia}, {Leclerc}, {Lecoeur-Taibi},
  {Lenhardt}, {Leroux}, {Liao}, {Licata}, {Lindstr{\o}m}, {Lister}, {Livanou},
  {Lobel}, {L{\'o}pez}, {Managau}, {Mann}, {Mantelet}, {Marchal}, {Marchant},
  {Marconi}, {Marinoni}, {Marschalk{\'o}}, {Marshall}, {Martino}, {Marton},
  {Mary}, {Massari}, {Matijevi{\v{c}}}, {Mazeh}, {McMillan}, {Messina},
  {Michalik}, {Millar}, {Molina}, {Molinaro}, {Moln{\'a}r}, {Montegriffo},
  {Mor}, {Morbidelli}, {Morel}, {Morris}, {Mulone}, {Muraveva}, {Musella},
  {Nelemans}, {Nicastro}, {Noval}, {O'Mullane}, {Ord{\'e}novic},
  {Ord{\'o}{\~n}ez-Blanco}, {Osborne}, {Pagani}, {Pagano}, {Pailler},
  {Palacin}, {Palaversa}, {Panahi}, {Pawlak}, {Piersimoni}, {Pineau}, {Plachy},
  {Plum}, {Poggio}, {Poujoulet}, {Pr{\v{s}}a}, {Pulone}, {Racero}, {Ragaini},
  {Rambaux}, {Ramos-Lerate}, {Regibo}, {Reyl{\'e}}, {Riclet}, {Ripepi}, {Riva},
  {Rivard}, {Rixon}, {Roegiers}, {Roelens}, {Romero-G{\'o}mez}, {Rowell},
  {Royer}, {Ruiz-Dern}, {Sadowski}, {Sagrist{\`a} Sell{\'e}s}, {Sahlmann},
  {Salgado}, {Salguero}, {Sanna}, {Santana-Ros}, {Sarasso}, {Savietto},
  {Schultheis}, {Sciacca}, {Segol}, {Segovia}, {S{\'e}gransan}, {Shih},
  {Siltala}, {Silva}, {Smart}, {Smith}, {Solano}, {Solitro}, {Sordo}, {Soria
  Nieto}, {Souchay}, {Spagna}, {Spoto}, {Stampa}, {Steele},
  {Steidelm{\"u}ller}, {Stephenson}, {Stoev}, {Suess}, {Surdej}, {Szabados},
  {Szegedi-Elek}, {Tapiador}, {Taris}, {Tauran}, {Taylor}, {Teixeira},
  {Terrett}, {Teyssandier}, {Thuillot}, {Titarenko}, {Torra Clotet}, {Turon},
  {Ulla}, {Utrilla}, {Uzzi}, {Vaillant}, {Valentini}, {Valette}, {van Elteren},
  {Van Hemelryck}, {van Leeuwen}, {Vaschetto}, {Vecchiato}, {Veljanoski},
  {Viala}, {Vicente}, {Vogt}, {von Essen}, {Voss}, {Votruba}, {Voutsinas},
  {Walmsley}, {Weiler}, {Wertz}, {Wevers}, {Wyrzykowski}, {Yoldas},
  {{\v{Z}}erjal}, {Ziaeepour}, {Zorec}, {Zschocke}, {Zucker}, {Zurbach}, \&
  {Zwitter}}]{GaiaDR2}
{Gaia Collaboration}, {Brown}, A.~G.~A., {Vallenari}, A., {et~al.} 2018, \aap,
  616, A1

\bibitem[{{Garc{\'{\i}}a} {et~al.}(2014){Garc{\'{\i}}a}, {Mathur}, {Pires},
  {R{\'e}gulo}, {Bellamy}, {Pall{\'e}}, {Ballot}, {Barcel{\'o} Forteza},
  {Beck}, {Bedding}, {Ceillier}, {Roca Cort{\'e}s}, {Salabert}, \&
  {Stello}}]{Garcia2014}
{Garc{\'{\i}}a}, R.~A., {Mathur}, S., {Pires}, S., {et~al.} 2014, A\&A, 568,
  A10

\bibitem[{{Gaulme} \& {Guzik}(2019)}]{GaulmeGuzik2019}
{Gaulme}, P. \& {Guzik}, J.~A. 2019, \aap, 630, A106

\bibitem[{{Gaulme} {et~al.}(2014){Gaulme}, {Jackiewicz}, {Appourchaux}, \&
  {Mosser}}]{Gaulme2014}
{Gaulme}, P., {Jackiewicz}, J., {Appourchaux}, T., \& {Mosser}, B. 2014, ApJ,
  785, 5

\bibitem[{{Gaulme} {et~al.}(2020){Gaulme}, {Jackiewicz}, {Spada}, {Chojnowski},
  {Mosser}, {McKeever}, {Hedlund}, {Vrard}, {Benbakoura}, \&
  {Damiani}}]{Gaulme2020}
{Gaulme}, P., {Jackiewicz}, J., {Spada}, F., {et~al.} 2020, \aap, 639, A63

\bibitem[{{Gaulme} {et~al.}(2016){Gaulme}, {McKeever}, {Jackiewicz}, {Rawls},
  {Corsaro}, {Mosser}, {Southworth}, {Mahadevan}, {Bender}, \&
  {Deshpande}}]{Gaulme2016}
{Gaulme}, P., {McKeever}, J., {Jackiewicz}, J., {et~al.} 2016, \apj, 832, 121

\bibitem[{{Gaulme} {et~al.}(2013){Gaulme}, {McKeever}, {Rawls}, {Jackiewicz},
  {Mosser}, \& {Guzik}}]{Gaulme2013}
{Gaulme}, P., {McKeever}, J., {Rawls}, M.~L., {et~al.} 2013, ApJ, 767, 82

\bibitem[{{Han} {et~al.}(2002){Han}, {Podsiadlowski}, {Maxted}, {Marsh}, \&
  {Ivanova}}]{han2002}
{Han}, Z., {Podsiadlowski}, P., {Maxted}, P.~F.~L., {Marsh}, T.~R., \&
  {Ivanova}, N. 2002, MNRAS, 336, 449

\bibitem[{{Handberg} {et~al.}(2017){Handberg}, {Brogaard}, {Miglio}, {Bossini},
  {Elsworth}, {Slumstrup}, {Davies}, \& {Chaplin}}]{Handberg2017}
{Handberg}, R., {Brogaard}, K., {Miglio}, A., {et~al.} 2017, \mnras, 472, 979

\bibitem[{Harris {et~al.}(2020)Harris, Millman, van~der Walt, Gommers,
  Virtanen, Cournapeau, Wieser, Taylor, Berg, Smith, \& et~al.}]{Harris_2020}
Harris, C.~R., Millman, K.~J., van~der Walt, S.~J., {et~al.} 2020, Nature, 585,
  357–362

\bibitem[{{Hon} {et~al.}(2021){Hon}, {Huber}, {Kuszlewicz}, {Stello}, {Sharma},
  {Tayar}, {Zinn}, {Vrard}, \& {Pinsonneault}}]{Hon2021}
{Hon}, M., {Huber}, D., {Kuszlewicz}, J.~S., {et~al.} 2021, \apj, 919, 131

\bibitem[{{Howell} {et~al.}(2014){Howell}, {Sobeck}, {Haas}, {Still},
  {Barclay}, {Mullally}, {Troeltzsch}, {Aigrain}, {Bryson}, {Caldwell},
  {Chaplin}, {Cochran}, {Huber}, {Marcy}, {Miglio}, {Najita}, {Smith},
  {Twicken}, \& {Fortney}}]{Howell2014}
{Howell}, S.~B., {Sobeck}, C., {Haas}, M., {et~al.} 2014, \pasp, 126, 398

\bibitem[{{Hunter}(2007)}]{4160265}
{Hunter}, J.~D. 2007, Computing in Science Engineering, 9, 90

\bibitem[{{Hut}(1980)}]{Hut1980}
{Hut}, P. 1980, \aap, 92, 167

\bibitem[{{Jackiewicz}(2021)}]{Jackiewicz2021}
{Jackiewicz}, J. 2021, Frontiers in Astronomy and Space Sciences, 7, 102

\bibitem[{{Johnston} {et~al.}(2019){Johnston}, {Pavlovski}, \&
  {Tkachenko}}]{Johnston2019}
{Johnston}, C., {Pavlovski}, K., \& {Tkachenko}, A. 2019, \aap, 628, A25

\bibitem[{{Kallinger} {et~al.}(2019){Kallinger}, {Beck}, {Hekker}, {Huber},
  {Kuschnig}, {Rockenbauer}, {Winter}, {Weiss}, {Handler}, {Moffat},
  {Pigulski}, {Popowicz}, {Wade}, \& {Zwintz}}]{Kallinger2019}
{Kallinger}, T., {Beck}, P.~G., {Hekker}, S., {et~al.} 2019, \aap, 624, A35

\bibitem[{{Kallinger} {et~al.}(2018){Kallinger}, {Beck}, {Stello}, \&
  {Garcia}}]{Kallinger2018}
{Kallinger}, T., {Beck}, P.~G., {Stello}, D., \& {Garcia}, R.~A. 2018, \aap,
  616, A104

\bibitem[{{Kirk} {et~al.}(2016){Kirk}, {Conroy}, {Pr{\v s}a}, {Abdul-Masih},
  {Kochoska}, {Matijevi{\v c}}, {Hambleton}, {Barclay}, {Bloemen}, {Boyajian},
  {Doyle}, {Fulton}, {Hoekstra}, {Jek}, {Kane}, {Kostov}, {Latham}, {Mazeh},
  {Orosz}, {Pepper}, {Quarles}, {Ragozzine}, {Shporer}, {Southworth},
  {Stassun}, {Thompson}, {Welsh}, {Agol}, {Derekas}, {Devor}, {Fischer},
  {Green}, {Gropp}, {Jacobs}, {Johnston}, {LaCourse}, {Saetre}, {Schwengeler},
  {Toczyski}, {Werner}, {Garrett}, {Gore}, {Martinez}, {Spitzer}, {Stevick},
  {Thomadis}, {Halley Vrijmoet}, {Yenawine}, {Batalha}, \&
  {Borucki}}]{Kirk2016}
{Kirk}, B., {Conroy}, K., {Pr{\v s}a}, A., {et~al.} 2016, \aj, 151, 68

\bibitem[{{Kjeldsen} \& {Bedding}(1995)}]{Kjeldsen1995}
{Kjeldsen}, H. \& {Bedding}, T.~R. 1995, A\&A, 293, 87

\bibitem[{{Kumar} {et~al.}(1995){Kumar}, {Ao}, \& {Quataert}}]{kumar1995}
{Kumar}, P., {Ao}, C.~O., \& {Quataert}, E.~J. 1995, ApJ, 449, 294

\bibitem[{{Li} {et~al.}(2018){Li}, {Bedding}, {Huber}, {Ball}, {Stello},
  {Murphy}, \& {Bland-Hawthorn}}]{LiT2018}
{Li}, T., {Bedding}, T.~R., {Huber}, D., {et~al.} 2018, \mnras, 475, 981

\bibitem[{{Mackereth} {et~al.}(2021){Mackereth}, {Miglio}, {Elsworth},
  {Mosser}, {Mathur}, {Garcia}, {Nardiello}, {Hall}, {Vrard}, {Ball}, {Basu},
  {Beaton}, {Beck}, {Bergemann}, {Bossini}, {Casagrande}, {Campante},
  {Chaplin}, {Chiappini}, {Girardi}, {J{\o}rgensen}, {Khan}, {Montalb{\'a}n},
  {Nielsen}, {Pinsonneault}, {Rodrigues}, {Serenelli}, {Silva Aguirre},
  {Stello}, {Tayar}, {Teske}, {van Saders}, \& {Willett}}]{Mackereth2021}
{Mackereth}, J.~T., {Miglio}, A., {Elsworth}, Y., {et~al.} 2021, \mnras, 502,
  1947

\bibitem[{{Mathur} {et~al.}(2019){Mathur}, {Garc{\'\i}a}, {Bugnet}, {Santos},
  {Santiago}, \& {Beck}}]{Mathur2019}
{Mathur}, S., {Garc{\'\i}a}, R.~A., {Bugnet}, L., {et~al.} 2019, FrASS, 6, 46

\bibitem[{{Mathur} {et~al.}(2010){Mathur}, {Garc{\'{\i}}a}, {R{\'e}gulo},
  {Creevey}, {Ballot}, {Salabert}, {Arentoft}, {Quirion}, {Chaplin}, \&
  {Kjeldsen}}]{Mathur2010}
{Mathur}, S., {Garc{\'{\i}}a}, R.~A., {R{\'e}gulo}, C., {et~al.} 2010, \aap,
  511, A46

\bibitem[{{Merc} {et~al.}(2021){Merc}, {Kalup}, {Rathour}, {S{\'a}nchez Arias},
  \& {Beck}}]{Merc2021}
{Merc}, J., {Kalup}, C., {Rathour}, R.~S., {S{\'a}nchez Arias}, J.~P., \&
  {Beck}, P.~G. 2021, Contributions of the Astronomical Observatory Skalnate
  Pleso, 51, 45

\bibitem[{{Milliman} {et~al.}(2014){Milliman}, {Mathieu}, {Geller}, {Gosnell},
  {Meibom}, \& {Platais}}]{Milliman2014}
{Milliman}, K.~E., {Mathieu}, R.~D., {Geller}, A.~M., {et~al.} 2014, \aj, 148,
  38

\bibitem[{{Moe} \& {Di Stefano}(2017)}]{MoeStefano2017}
{Moe}, M. \& {Di Stefano}, R. 2017, \apjs, 230, 15

\bibitem[{{Mosser} {et~al.}(2013){Mosser}, {Michel}, {Belkacem}, {Goupil},
  {Baglin}, {Barban}, {Provost}, {Samadi}, {Auvergne}, \&
  {Catala}}]{Mosser2013}
{Mosser}, B., {Michel}, E., {Belkacem}, K., {et~al.} 2013, A\&A, 550, A126

\bibitem[{Oliphant(2006)}]{numpy}
Oliphant, T. 2006, {NumPy}: A guide to {NumPy}, USA: Trelgol Publishing

\bibitem[{{Pires} {et~al.}(2015){Pires}, {Mathur}, {Garc{\'\i}a}, {Ballot},
  {Stello}, \& {Sato}}]{Pires2015}
{Pires}, S., {Mathur}, S., {Garc{\'\i}a}, R.~A., {et~al.} 2015, \aap, 574, A18

\bibitem[{{Pourbaix} {et~al.}(2004){Pourbaix}, {Tokovinin}, {Batten}, {Fekel},
  {Hartkopf}, {Levato}, {Morrell}, {Torres}, \& {Udry}}]{Pourbaix2004}
{Pourbaix}, D., {Tokovinin}, A.~A., {Batten}, A.~H., {et~al.} 2004, \aap, 424,
  727

\bibitem[{{Pr{\v s}a} {et~al.}(2011){Pr{\v s}a}, {Batalha}, {Slawson}, {Doyle},
  {Welsh}, {Orosz}, {Seager}, {Rucker}, {Mjaseth}, {Engle}, {Conroy},
  {Jenkins}, {Caldwell}, {Koch}, \& {Borucki}}]{Prsa2011}
{Pr{\v s}a}, A., {Batalha}, N., {Slawson}, R.~W., {et~al.} 2011, AJ, 141, 83

\bibitem[{{Pr{\v{s}}a}(2018)}]{Prsa2018}
{Pr{\v{s}}a}, A. 2018, {\textsc{Phoebe 2} - Modeling and Analysis of Eclipsing
  Binary Stars}

\bibitem[{{Rauer} {et~al.}(2014){Rauer}, {Catala}, {Aerts}, {Appourchaux},
  {Benz}, {Brandeker}, {Christensen-Dalsgaard}, {Deleuil}, {Gizon}, {Goupil},
  {G{\"u}del}, {Janot-Pacheco}, {Mas-Hesse}, {Pagano}, {Piotto}, {Pollacco},
  {Santos}, {Smith}, {Su{\'a}rez}, {Szab{\'o}}, {Udry}, {Adibekyan}, {Alibert},
  {Almenara}, {Amaro-Seoane}, {Eiff}, {Asplund}, {Antonello}, {Barnes},
  {Baudin}, {Belkacem}, {Bergemann}, {Bihain}, {Birch}, {Bonfils}, {Boisse},
  {Bonomo}, {Borsa}, {Brand{\~a}o}, {Brocato}, {Brun}, {Burleigh}, {Burston},
  {Cabrera}, {Cassisi}, {Chaplin}, {Charpinet}, {Chiappini}, {Church},
  {Csizmadia}, {Cunha}, {Damasso}, {Davies}, {Deeg}, {D{\'{\i}}az}, {Dreizler},
  {Dreyer}, {Eggenberger}, {Ehrenreich}, {Eigm{\"u}ller}, {Erikson}, {Farmer},
  {Feltzing}, {de Oliveira Fialho}, {Figueira}, {Forveille}, {Fridlund},
  {Garc{\'{\i}}a}, {Giommi}, {Giuffrida}, {Godolt}, {Gomes da Silva},
  {Granzer}, {Grenfell}, {Grotsch-Noels}, {G{\"u}nther}, {Haswell}, {Hatzes},
  {H{\'e}brard}, {Hekker}, {Helled}, {Heng}, {Jenkins}, {Johansen},
  {Khodachenko}, {Kislyakova}, {Kley}, {Kolb}, {Krivova}, {Kupka}, {Lammer},
  {Lanza}, {Lebreton}, {Magrin}, {Marcos-Arenal}, {Marrese}, {Marques},
  {Martins}, {Mathis}, {Mathur}, {Messina}, {Miglio}, {Montalban}, {Montalto},
  {Monteiro}, {Moradi}, {Moravveji}, {Mordasini}, {Morel}, {Mortier},
  {Nascimbeni}, {Nelson}, {Nielsen}, {Noack}, {Norton}, {Ofir}, {Oshagh},
  {Ouazzani}, {P{\'a}pics}, {Parro}, {Petit}, {Plez}, {Poretti}, {Quirrenbach},
  {Ragazzoni}, {Raimondo}, {Rainer}, {Reese}, {Redmer}, {Reffert},
  {Rojas-Ayala}, {Roxburgh}, {Salmon}, {Santerne}, {Schneider}, {Schou},
  {Schuh}, {Schunker}, {Silva-Valio}, {Silvotti}, {Skillen}, {Snellen}, {Sohl},
  {Sousa}, {Sozzetti}, {Stello}, {Strassmeier}, {{\v S}vanda}, {Szab{\'o}},
  {Tkachenko}, {Valencia}, {Van Grootel}, {Vauclair}, {Ventura}, {Wagner},
  {Walton}, {Weingrill}, {Werner}, {Wheatley}, \& {Zwintz}}]{Rauer2014}
{Rauer}, H., {Catala}, C., {Aerts}, C., {et~al.} 2014, Experimental Astronomy,
  38, 249

\bibitem[{{Rawls} {et~al.}(2016){Rawls}, {Gaulme}, {McKeever}, {Jackiewicz},
  {Orosz}, {Corsaro}, {Beck}, {Mosser}, {Latham}, \& {Latham}}]{Rawls2016}
{Rawls}, M.~L., {Gaulme}, P., {McKeever}, J., {et~al.} 2016, \apj, 818, 108

\bibitem[{{Ricker} {et~al.}(2014){Ricker}, {Winn}, {Vanderspek}, {Latham},
  {Bakos}, {Bean}, {Berta-Thompson}, {Brown}, {Buchhave}, {Butler}, {Butler},
  {Chaplin}, {Charbonneau}, {Christensen-Dalsgaard}, {Clampin}, {Deming},
  {Doty}, {De Lee}, {Dressing}, {Dunham}, {Endl}, {Fressin}, {Ge}, {Henning},
  {Holman}, {Howard}, {Ida}, {Jenkins}, {Jernigan}, {Johnson}, {Kaltenegger},
  {Kawai}, {Kjeldsen}, {Laughlin}, {Levine}, {Lin}, {Lissauer}, {MacQueen},
  {Marcy}, {McCullough}, {Morton}, {Narita}, {Paegert}, {Palle}, {Pepe},
  {Pepper}, {Quirrenbach}, {Rinehart}, {Sasselov}, {Sato}, {Seager},
  {Sozzetti}, {Stassun}, {Sullivan}, {Szentgyorgyi}, {Torres}, {Udry}, \&
  {Villasenor}}]{Ricker2014}
{Ricker}, G.~R., {Winn}, J.~N., {Vanderspek}, R., {et~al.} 2014, in SPIE 914320

\bibitem[{{Silva Aguirre} {et~al.}(2020){Silva Aguirre}, {Stello}, {Stokholm},
  {Mosumgaard}, {Ball}, {Basu}, {Bossini}, {Bugnet}, {Buzasi}, {Campante},
  {Carboneau}, {Chaplin}, {Corsaro}, {Davies}, {Elsworth}, {Garc{\'\i}a},
  {Gaulme}, {Hall}, {Handberg}, {Hon}, {Kallinger}, {Kang}, {Lund}, {Mathur},
  {Mints}, {Mosser}, {{\c{C}}elik Orhan}, {Rodrigues}, {Vrard}, {Y{\i}ld{\i}z},
  {Zinn}, {{\"O}rtel}, {Beck}, {Bell}, {Guo}, {Jiang}, {Kuszlewicz}, {Kuehn},
  {Li}, {Lundkvist}, {Pinsonneault}, {Tayar}, {Cunha}, {Hekker}, {Huber},
  {Miglio}, {F.~G. Monteiro}, {Slumstrup}, {Winther}, {Angelou}, {Benomar},
  {B{\'o}di}, {De Moura}, {Deheuvels}, {Derekas}, {Di Mauro}, {Dupret},
  {Jim{\'e}nez}, {Lebreton}, {Matthews}, {Nardetto}, {do Nascimento},
  {Pereira}, {Rodr{\'\i}guez D{\'\i}az}, {Serenelli}, {Spitoni},
  {Stonkut{\.{e}}}, {Su{\'a}rez}, {Szab{\'o}}, {Van Eylen}, {Ventura}, {Verma},
  {Weiss}, {Wu}, {Barclay}, {Christensen-Dalsgaard}, {Jenkins}, {Kjeldsen},
  {Ricker}, {Seager}, \& {Vanderspek}}]{SilvaAguirre2020}
{Silva Aguirre}, V., {Stello}, D., {Stokholm}, A., {et~al.} 2020, \apjl, 889,
  L34

\bibitem[{{Stello} {et~al.}(2011){Stello}, {Meibom}, {Gilliland}, {Grundahl},
  {Hekker}, {Mosser}, {Kallinger}, {Mathur}, {Garc{\'\i}a}, {Huber}, {Basu},
  {Bedding}, {Brogaard}, {Chaplin}, {Elsworth}, {Molenda-{\.Z}akowicz},
  {Szab{\'o}}, {Still}, {Jenkins}, {Christensen-Dalsgaard}, {Kjeldsen},
  {Serenelli}, \& {Wohler}}]{Stello2011Clusters}
{Stello}, D., {Meibom}, S., {Gilliland}, R.~L., {et~al.} 2011, \apj, 739, 13

\bibitem[{{Tassoul}(1980)}]{Tassoul1980}
{Tassoul}, M. 1980, ApJS, 43, 469

\bibitem[{{Theme{\ss}l} {et~al.}(2018){Theme{\ss}l}, {Hekker}, {Southworth},
  {Beck}, {Pavlovski}, {Tkachenko}, {Angelou}, {Ball}, {Barban}, {Corsaro},
  {Elsworth}, {Handberg}, \& {Kallinger}}]{Themessl2018}
{Theme{\ss}l}, N., {Hekker}, S., {Southworth}, J., {et~al.} 2018, \mnras

\bibitem[{{Thompson} {et~al.}(2012){Thompson}, {Everett}, {Mullally},
  {Barclay}, {Howell}, {Still}, {Rowe}, {Christiansen}, {Kurtz}, {Hambleton},
  {Twicken}, {Ibrahim}, \& {Clarke}}]{Thompson2012}
{Thompson}, S.~E., {Everett}, M., {Mullally}, F., {et~al.} 2012, ApJ, 753, 86

\bibitem[{{Torres} {et~al.}(2010){Torres}, {Andersen}, \&
  {Gim{\'e}nez}}]{Torres2010AccurateMasses}
{Torres}, G., {Andersen}, J., \& {Gim{\'e}nez}, A. 2010, A\&Ar, 18, 67

\bibitem[{{van Leeuwen}(2007)}]{vanLeeuwen2007}
{van Leeuwen}, F. 2007, A\&A, 474, 653

\bibitem[{Van~Rossum \& Drake(2009)}]{10.5555/1593511}
Van~Rossum, G. \& Drake, F.~L. 2009, Python 3 Reference Manual (Scotts Valley,
  CA: CreateSpace)

\bibitem[{{Virtanen} {et~al.}(2020){Virtanen}, {Gommers}, {Oliphant},
  {Haberland}, {Reddy}, {Cournapeau}, {Burovski}, {Peterson}, {Weckesser},
  {Bright}, {van der Walt}, {Brett}, {Wilson}, {Jarrod Millman}, {Mayorov},
  {Nelson}, {Jones}, {Kern}, {Larson}, {Carey}, {Polat}, {Feng}, {Moore}, {Vand
  erPlas}, {Laxalde}, {Perktold}, {Cimrman}, {Henriksen}, {Quintero}, {Harris},
  {Archibald}, {Ribeiro}, {Pedregosa}, {van Mulbregt}, \&
  {Contributors}}]{2020SciPy-NMeth}
{Virtanen}, P., {Gommers}, R., {Oliphant}, T.~E., {et~al.} 2020, Nature
  Methods, 17, 261

\bibitem[{{Weiss} {et~al.}(2014){Weiss}, {Rucinski}, {Moffat},
  {Schwarzenberg-Czerny}, {Koudelka}, {Grant}, {Zee}, {Kuschnig}, {Mochnacki},
  {Matthews}, {Orleanski}, {Pamyatnykh}, {Pigulski}, {Alves}, {Guedel},
  {Handler}, {Wade}, \& {Zwintz}}]{Weiss2014}
{Weiss}, W.~W., {Rucinski}, S.~M., {Moffat}, A.~F.~J., {et~al.} 2014, \pasp,
  126, 573

\bibitem[{{White} {et~al.}(2013){White}, {Huber}, {Maestro}, {Bedding},
  {Ireland}, {Baron}, {Boyajian}, {Che}, {Monnier}, {Pope}, {Roettenbacher},
  {Stello}, {Tuthill}, {Farrington}, {Goldfinger}, {McAlister}, {Schaefer},
  {Sturmann}, {Sturmann}, {ten Brummelaar}, \& {Turner}}]{White2013}
{White}, T.~R., {Huber}, D., {Maestro}, V., {et~al.} 2013, \mnras, 433, 1262

\bibitem[{{Yu} {et~al.}(2018){Yu}, {Huber}, {Bedding}, {Stello}, {Hon},
  {Murphy}, \& {Khanna}}]{Yu2018}
{Yu}, J., {Huber}, D., {Bedding}, T.~R., {et~al.} 2018, \apjs, 236, 42

\bibitem[{{Zahn}(2013)}]{Zahn2013}
{Zahn}, J.-P. 2013, in Lecture Notes in Physics, Vol. 861, 301

\end{thebibliography}

\begin{appendix}

\section{Catalogue of oscillating RG binaries in the literature \label{appendix:publishedSample}}
In the scientific literature, currently 83 systems with an oscillating RG star are known. The Tables\,\ref{tab:publishedSampleComplete}, \ref{tab:publishedSampleIncomplete}, and \ref{tab:sampleHardlyAnyInformation} cite the papers in which a system has been seismically analysed, using the following abbreviations. The references for the quoted values is given first, followed by other references for this star in the bracket:
\citet[B14]{Beck2014a}, 
\citet[B15a]{Beck2015a},
\citet[B15b]{Beck2015Toulouse},
\citet[B18]{Beck2018Asterix},
\citet[F13]{Frandsen2013},
\citet[G13]{Gaulme2013},
\citet[G14]{Gaulme2014},
\citet[G16]{Gaulme2016},
\citet[R16]{Rawls2016},
%\citet[B16]{Brogaard2016},
\citet[L18]{LiT2018},
\citet[B18]{Brogaard2018},
\citet[T18]{Themessl2018},
\citet[GG19]{GaulmeGuzik2019},
\citet[G20]{Gaulme2020},
\citet[M21]{Merc2021},
and \citet[B21]{Benbakoura2021}

In addition to the already described parameters, Tables\,\ref{tab:publishedSampleComplete} and \ref{tab:publishedSampleIncomplete} also report the evolutionary status, discriminating between RGB and RC RG primaries, the value of the asymptotic period spacing of gravity dipole modes, $\Delta\Pi_{1}$ and the mass ratio for each system, M$_2$/M$_1$.

\begin{sidewaystable*}
\caption{Red-giant binary systems with determined seismic and orbital parameters.  \label{tab:publishedSampleComplete}}
    \centering
\tabcolsep=2pt
    \begin{tabular}{rrr | rllll | lll | l}
    \hline\hline
\multicolumn{1}{c}{KIC}                         &       
\multicolumn{1}{c}{Typ}                         &
\multicolumn{1}{c}{Evol.}                               &       
\multicolumn{1}{c}{V}                                   &       
\multicolumn{1}{c}{T$_\mathrm{eff}$}    &       
\multicolumn{1}{c}{$\nu_\mathrm{max}$}  &       
\multicolumn{1}{c}{$\Delta\nu$}                 &       
\multicolumn{1}{c}{$\Delta\Pi_{1}$}                     &       
\multicolumn{1}{c}{P$_\mathrm{orb}$}    &       
\multicolumn{1}{c}{$e$}                         &       
\multicolumn{1}{c}{M$_2$/M$_1$}         &       
\multicolumn{1}{l}{Literature References}       \\
                                &       
                        &
\multicolumn{1}{c}{State}                       &       
\multicolumn{1}{c}{[mag]}                       &       
\multicolumn{1}{c}{[K]}                 &       
\multicolumn{1}{c}{[$\mu$Hz]}           &       
\multicolumn{1}{c}{[$\mu$Hz]}           &       
\multicolumn{1}{c}{[s]}                 &       
\multicolumn{1}{c}{[d]}                 &       
                         \\[2mm] \hline

9163796A        &       HB. SB2. PB2    &       RGB     &       9.8     &       5020    $\pm$   100     &       165.3   $\pm$   1.3     &       12.85   $\pm$   0.03    &               &       121.30  $\pm$   0.01    &       0.692   $\pm$   0.002   &       0.99    $\pm$   0.005   &       B18     (B14, GG19)   \\
9163796B        &               &       RGB     &               &       5650    $\pm$   70      &       410     $\pm$   50      &                               &               &                               &                               &                               &                       \\
9246715A        &       EC. SB2. PB2    &       RC      &       9.7     &       4930    $\pm$   230     &       106.4   $\pm$   0.8     &       8.3     $\pm$   0.02    &               &       171.27688       $\pm$   0.00001 &       0.3559  $\pm$   0.0003  &       0.90    $\pm$   0.001   &       R16     (G13, G14, GG19)      \\
9246715B        &               &       RC      &               &       4930    $\pm$   230     &       106.4   $\pm$   0.8     &                               &               &                               &                               &                               &                       \\ \hline
2444348 &       HB      &       RGB     &       10.7    &       4565                    &       30.5    $\pm$   0.3     &       3.26    $\pm$   0.01    &               &       103.50  $\pm$   0.01    &       0.48    $\pm$   0.01    &                               &       B14     (GG19)  \\
2697935 &       EC. HB  &       RGB     &       11.0    &       4883                    &       405.6   $\pm$   3       &       28                      &               &       21.50   $\pm$   0.01    &       0.41    $\pm$   0.02    &                               &       B14     (GG19)  \\
2720096 &       HB      &       RGB     &               &       4812                    &       110.1   $\pm$   0.7     &       9.17    $\pm$   0.01    &               &       26.70   $\pm$   0.01    &       0.49    $\pm$   0.01    &                               &       B14     (GG19)  \\
4054905 &       SB2. EC &       RC      &               &       4790    $\pm$   190     &       48.13   $\pm$   0.21    &       5.42    $\pm$   0.01    &       159.5   &       274.7306        $\pm$   0.0004  &       0.372   $\pm$   0.002   &       0.98                    &       B21     (GG19)  \\
4360072 &       EC      &       RC      &       11.4    &       5020    $\pm$   210     &       31.81   $\pm$   0.06    &       3.90    $\pm$   0.01    &       392.2   &       1084.76 $\pm$   0.01    &       0.152   $\pm$   0.001   &                               &       B21     (GG19)  \\
4663185 &       EC. SB2 &       RGB     &       11.5    &       4803    $\pm$   91      &       54.09   $\pm$   0.24    &       5.212   $\pm$   0.019   &               &       358.09  $\pm$   0.0003  &       0.43    $\pm$   0.01    &       0.99    $\pm$   0.08    &       G16     (G13)   \\
4663623 &       SB2. EC &       RC      &               &       4812    $\pm$   92      &       54.09   $\pm$   0.24    &       5.21    $\pm$   0.02    &       363.6   &       358.0900        $\pm$   0.0001  &       0.399   $\pm$   0.001   &       0.96    $\pm$           &       B21     (G14, G16, B18, L18, GG19)    \\
5006817 &       HB      &       RGB     &       11.2    &       5000                    &       145.9   $\pm$   0.5     &       11.64   $\pm$   0.01    &               &       94.812  $\pm$   0.002   &       0.7069  $\pm$   0.0002  &                               &       B14     (GG19, M21)    \\
5039392 &       HB      &       RGB     &               &       4110                    &       6.2     $\pm$   0.1     &       1.13    $\pm$   0.01    &               &       236.70  $\pm$   0.02    &       0.44    $\pm$   0.01    &                               &       B14     (GG19)  \\
5179609 &       EC. SB2 &       RGB     &               &       4887    $\pm$   91      &       321.84  $\pm$   1       &       22.21   $\pm$   0.05    &               &       43.931080       $\pm$   0.00002 &       0.15    $\pm$   0.001   &       0.51    $\pm$   0.02    &       G16     (G13, G14, B21, GG19) \\
5308778 &       EC. SB2 &       RGB     &               &       5044    $\pm$   91      &       48.47   $\pm$   1.1     &       5.05    $\pm$   0.05    &               &       40.5661 $\pm$   0.0003  &       0.006   $\pm$   0.005   &       0.43    $\pm$   0.03    &       G16     (G13, G14, B21, GG19) \\
5640750 &       EC      &       RGB     &       11.9    &       4525    $\pm$   75      &       24.1    $\pm$   0.2     &       2.969   $\pm$   0.006   &               &       987.398 $\pm$   0.006   &       0.322   $\pm$   0.008   &       0.85    $\pm$   0.03    &       T18     (G13, G14, G16, B21, T18, GG19)       \\
5786154 &       EC. SB2 &       RGB     &               &       4747    $\pm$   100     &       29.75   $\pm$   0.16    &       3.52    $\pm$   0.01    &               &       197.9180        $\pm$   0.0004  &       0.3764  $\pm$   0.0009  &       0.96    $\pm$   0.07    &       G16     (G13, G14, L18, B21, GG19)    \\
5866138 &       EC      &       RC      &       11.5    &       4960    $\pm$   120     &       83.71   $\pm$   0.46    &       7.25    $\pm$   0.01    &       272.4   &       342.259 $\pm$   0.008   &       0.7158  $\pm$   0.0004  &                               &       B21     (GG19)  \\
6757558 &       EC      &       RGB     &               &       4590    $\pm$   110     &       129.38  $\pm$   0.28    &       11.28   $\pm$   0.01    &       80.1    &       421.190 $\pm$   0.009   &       0.22    $\pm$   0.03    &                               &       B21     (GG19)  \\
7037405 &       EC. SB2 &       RGB     &               &       4516    $\pm$   36      &       21.75   $\pm$   0.14    &       2.79    $\pm$   0.01    &               &       207.1083        $\pm$   0.0007  &       0.238   $\pm$   0.004   &       0.91    $\pm$   0.04    &       G16     (G13, G14, L18, B18, B21, GG19)       \\
7293054 &       SB2. EC &       ?       &       11.6    &       4790    $\pm$   160     &       42.58   $\pm$   0.27    &       4.32    $\pm$   0.01    &               &       671.806 $\pm$   0.003   &       0.8     $\pm$   0.01    &       0.84    $\pm$           &       B21     (GG19)  \\
7377422 &       EC. SB2 &       RGB     &               &       4938    $\pm$   110     &       40.1    $\pm$   2.1     &       4.64    $\pm$   0.05    &               &       107.6213        $\pm$   0.0004  &       0.4377  $\pm$   0.0005  &       0.81    $\pm$   0.07    &       G16     (G14, GG19)   \\
7768447 &       EC      &       ?       &       12.0    &       4760    $\pm$   160     &       57.79   $\pm$   0.52    &       5.89    $\pm$   0.02    &               &       122.32  $\pm$   0.04    &       0.322   $\pm$   0.009   &                               &       B21     (GG19)  \\
8054233 &       EC. SB2 &       RGB     &               &       4971    $\pm$   90      &       46.49   $\pm$   0.33    &       4.81    $\pm$   0.015   &               &       1058.16 $\pm$   0.02    &       0.2718  $\pm$   0.004   &       0.69    $\pm$   0.04    &       G16     (G14, GG19)   \\
8095275 &       HB      &       RGB     &               &       4622                    &       69.3    $\pm$   0.3     &       6.81    $\pm$   0.01    &               &       23.00   $\pm$   0.01    &       0.32    $\pm$   0.01    &                               &       B14     (G13, B14, GG19)      \\
8144355 &       HB      &       RGB     &               &       4875                    &       179     $\pm$   2       &       13.95   $\pm$   0.04    &               &       80.55   $\pm$   0.01    &       0.76    $\pm$   0.01    &                               &       B14     (GG19)  \\
8210370 &       HB      &       RGB     &       11.4    &       4585                    &       44.1    $\pm$   0.8     &       4.69    $\pm$   0.02    &               &       153.50  $\pm$   0.01    &       0.7     $\pm$   0.01    &                               &       B14     (GG19)  \\
8410637 &       EC      &       RGB     &       11.3    &       4605    $\pm$   80      &       46.4    $\pm$   0.3     &       4.564   $\pm$   0.004   &               &       408.3248        $\pm$   0.0004  &       0.686   $\pm$   0.001   &       0.89    $\pm$   0.01    &       T18     (F13, G13, G14, L18, B21, GG19)       \\
8430105 &       EC. SB2 &       RGB     &       10.6    &       5042    $\pm$   68      &       76.7    $\pm$   0.57    &       7.14    $\pm$   0.031   &               &       63.32713        $\pm$   0.0003  &       0.2564  $\pm$   0.0002  &       0.63    $\pm$   0.01    &       G16     (G13, G14, B21, GG19) \\
8702921 &       EC. SB2 &       RGB     &               &       5058    $\pm$   86      &       195.57  $\pm$   0.47    &       14.07   $\pm$   0.01    &               &       19.38446        $\pm$   0.00002 &       0.0964  $\pm$   0.0008  &       0.16    $\pm$   0.01    &       G16     (G13, G14, B21, GG19) \\
8912308 &       HB      &       RGB     &       11.8    &       4872                    &       350     $\pm$   9       &       22.7                    &               &       20.17   $\pm$   0.01    &       0.23    $\pm$   0.01    &                               &       B14     (GG19)  \\
9151763 &       HB      &       RGB?    &               &       4290                    &       13.8    $\pm$   0.2     &       1.98    $\pm$   0.01    &               &       437.51  $\pm$   0.03    &       0.73    $\pm$   0.01    &                               &       B14     (GG19)  \\
9153621 &       SB2. EC &       ?       &               &       4760    $\pm$   190     &       38.22   $\pm$   0.3     &       4.28    $\pm$   0.01    &               &       305.792 $\pm$   0.005   &       0.7     $\pm$   0.003   &       0.86                    &       B21     (GG22)  \\
9408183 &       HB      &       RGB     &               &       4900                    &       164.8   $\pm$   0.2     &       13.29   $\pm$   0.02    &               &       49.70   $\pm$   0.01    &       0.42    $\pm$   0.01    &                               &       B14     (GG19)  \\
9540226 &       EC. SB2 &       RGB     &       12.4    &       4692    $\pm$   65      &       27.07   $\pm$   0.15    &       3.22    $\pm$   0.013   &               &       175.4439        $\pm$   0.0006  &       0.388   $\pm$   0.0002  &       0.74    $\pm$   0.04    &       G16     (G13, G14, B14, L18. B18. T18, GG19)  \\
9904059 &       EC      &       RGB     &               &       4830    $\pm$   160     &       140.61  $\pm$   0.45    &       11.91   $\pm$   0.01    &       79.8    &       102.963 $\pm$   0.001   &       0.32    $\pm$   0.01    &                               &       B21     (GG19)  \\
9970396 &       EC. SB2 &       RGB     &               &       4916    $\pm$   68      &       63.70   $\pm$   0.16    &       6.32    $\pm$   0.01    &               &       235.2985        $\pm$   0.0002  &       0.194   $\pm$   0.007   &       0.89    $\pm$   0.03    &       G16     (G13, G14, L18, B18, B21, GG19)       \\
10001167        &       EC. SB2 &       RGB     &       10.4    &       4700    $\pm$   66      &       19.90   $\pm$   0.09    &       2.762   $\pm$   0.012   &               &       120.3903        $\pm$   0.0005  &       0.159   $\pm$   0.003   &       0.98    $\pm$   0.07    &       G16     (G13, G14, GG19)      \\
10015516        &       EC      &       RC      &       10.9    &       4830    $\pm$   130     &       66.85   $\pm$   0.67    &       5.90    $\pm$   0.01    &       294.5   &       67.69217        $\pm$   0.00009 &       0       $\pm$   0.01    &                               &       B21     (GG19)  \\
10074700        &       EC      &       ?       &               &       5070    $\pm$   100     &       232     $\pm$   2       &       18.37   $\pm$   0.02    &               &       365.6340        $\pm$   0.0006  &       0.29    $\pm$   0.06    &                               &       B21     (GG19)  \\
10614012        &       EC. HB  &       RGB     &       10.0    &       4715                    &       70.2    $\pm$   0.9     &       6.54    $\pm$   0.02    &               &       132.13  $\pm$   0.01    &       0.71    $\pm$   0.01    &                               &       B14     (GG19)  \\\hline
\hline  \end{tabular}           
\end{sidewaystable*}

\begin{table*}%[t!]
\caption{Catalogue and parameters for oscillating RG binaries characterised through photometry from the \Kepler mission. \label{tab:publishedSampleIncomplete}}
    \centering
\tabcolsep=4.3pt
    \begin{tabular}{rrr | rllll | lll | l}
    \hline\hline
\multicolumn{1}{c}{KIC}                         &       
\multicolumn{1}{c}{Typ}                         &
\multicolumn{1}{c}{Evol.}                               &       
\multicolumn{1}{c}{V}                                   &       
\multicolumn{1}{c}{T$_\mathrm{eff}$}    &       
\multicolumn{1}{c}{$\nu_\mathrm{max}$}  &       
\multicolumn{1}{c}{$\Delta\nu$}                 &       
\multicolumn{1}{c}{$\Delta\Pi_{1}$}                     &       
\multicolumn{1}{c}{P$_\mathrm{orb}$}    &       
%\multicolumn{1}{c}{$e$}                                &       
%\multicolumn{1}{c}{M$_2$/M$_1$}                &       
\multicolumn{1}{l}{Literature}  \\
                        &       
                        &
\multicolumn{1}{c}{State}                       &       
\multicolumn{1}{c}{[mag]}                       &       
\multicolumn{1}{c}{[K]}                 &       
\multicolumn{1}{c}{[$\mu$Hz]}           &       
\multicolumn{1}{c}{[$\mu$Hz]}           &       
\multicolumn{1}{c}{[s]}                 &       
\multicolumn{1}{c}{[d]}                 &       
\multicolumn{1}{l}{References}  
                         \\[2mm] \hline

7431665 &       HB      &       RGB     &       11.3    &       4580                    &       54.0 $\pm$ 0.7       &       5.46 $\pm$ 0.02 &                   &   281.4           &       B14     (GG19)  \\
7799540 &       HB      &       RGB     &                   &   5177                    &       347.2 $\pm$ 5 &       24.0                        &               &   71.8            &       B14     (GG19)  \\
8803882 &       HB      &       RGB     &                   &   5043                    &       347.0 $\pm$ 3 &       22.6 $\pm$ 0.4  &                   &   89.7            &       B14     (GG19)  \\
11044668 &      HB      &       RGB?&   11.9    &       4565                    &       50.2 $\pm$ 0.2       &       5.56 $\pm$ 0.01 &                   &   139.5           &       B14     (GG19)  \\
8108336 &               &       ?       &                   &   4868 $\pm$ 170     &       46.6 $\pm$ 0.57 &       4.50                        &               &                            &       G20             \\
8515227 &               &       RGB     &       11.5    &       4778                    &       176 $\pm$ 2     &   14.60                   &       82.2    &                               &       G20             \\
9837673 &               &       RGB     &                   &   5062                    &       222 $\pm$ 2     &   15.13                   &       96.1    &                               &       G20             \\
10198347 &              &       RC      &       10.6    &       4924                    &       39.65 $\pm$ 0.50&     4.47                    &       276.6   &                               &       G20             \\
10811720 &              &       ?       &       12.2    &       4893                    &       38.98 $\pm$ 0.33&     4.37                    &                   &                           &       G20             \\
12314910 &              &       ?       &       11.6    &       4514                    &       23.96 $\pm$ 0.25&     2.97                    &                   &                           &       G20             \\
3458643 &               &       RC      &       10.5    &       5035                    &       70 $\pm$ 2     &   5.82                    &       213.9   &                               &       G20             \\
3967501 &               &       RC      &                   &   4682                    &       115 $\pm$ 8     &   9.95                    &       146.9   &                               &       G20             \\
4242873 &               &       RGB     &                   &   4848                    &       121.48 $\pm$ 0.89 &    10.56                   &       63.9    &                               &       G20             \\
5382824 &               &       RC      &       10.7    &       5114                    &       104.05 $\pm$ 0.81 &    7.90                        &   252.2   &                               &       G20             \\
5439339 &               &       RC      &       11.3    &       5098                    &       99 $\pm$ 1     &   7.86                    &       250.2   &                               &       G20             \\
6032639 &               &       RC      &       11.4    &       4862                    &       45.00 $\pm$ 0.68      &       4.77                    &       289.8   &                               &       G20             \\
6590195 &               &       RGB     &       12.5    &       4779                    &       113.11 $\pm$ 0.60 &    9.54                    &                   &                           &       G20             \\
6933666 &               &       ?       &                   &   4668                    &       32.60 $\pm$ 0.39      &       3.87            &                   &                           &       G20             \\
7103951 &               &       RC      &       9.6         &   4902                    &       52.70 $\pm$ 0.57      &       5.11            &       304.1   &                               &       G20             \\
7948193 &               &       RGB?&   11.1    &       4921                    &       113.70 $\pm$ 0.87&     9.09            &                   &                           &       G20             \\\hline
\end{tabular}   
%\tablefoot{asdf}
%\end{table*}

\vspace{25mm}

%\begin{table*}[t!]
\centering
\tabcolsep=15pt
\caption{Catalogue of RG binary system with limited information in the literature. \label{tab:sampleHardlyAnyInformation}}

\begin{tabular}{rrrrr}
\hline\hline
\multicolumn{1}{c}{KIC} &       
\multicolumn{1}{c}{Type}        &       
\multicolumn{1}{c}{T$_\mathrm{eff}$}    &       \multicolumn{1}{c}{P$_\mathrm{orbit}$}  &       Literature\\
&       &       
\multicolumn{1}{c}{[K]} &       
\multicolumn{1}{c}{[days]}      &
Reference\\ \hline
3851949     & EC    &   4981    &       54.77   &       GG19    \\
4078157     & EC    &   5547    &       16.02   &       GG19    \\
4769799     & EC    &   4911    &       21.93   &       GG19    \\
4999260     & EC    &   5048    &       0.38    &       GG19    \\
5877364     & EC    &           &       89.65   &       GG19    \\
6042191     & EC    &   4986    &       43.39   &       GG19    \\
6286155     & EC    &   5062    &       14.54   &       GG19    \\
6525209     & EC    &   5207    &       75.13   &       GG19    \\
6850665     & EC    &   4828    &       214.72  &       GG19    \\
8129189     & EC    &   5080    &       53.65   &       GG19    \\
8143170     & EC    &   4957    &       28.79   &       GG19    \\
8308347     & EC    &   4826    &       164.95  &       GG19    \\
8564976     & EC    &   4726    &       152.83  &       GG19    \\
9181877     & EC    &   4599    &       0.32    &       GG19    \\
10485250        & EC    &       4957    &       16.47   &       GG19    \\
10491544        & EC    &       4835    &       22.77   &       GG19    \\
10920813        & EC    &               &       53.74   &       GG19    \\
11135978        & EC    &       5004    &       0.29    &       GG19    \\
11768970        & EC    &       5038    &       15.54   &       GG19    \\
12367310        & EC    &       4965    &       8.63    &       GG19    \\ 
6042423     & HB    &   4689    &           &   B15b \\
10188415    & HB    &           &           &   B15b\\
10322133    & HB    &   5223    &           &   B15b \\
\hline
\end{tabular}
%\tablefoot{asdf}
\end{table*}

\section{Catalogue of new oscillating RG binaries \label{appendix:tables}}

Tables\,\ref{tab:targetsTESS}, \ref{tab:targetsKepler},
\ref{tab:BriteRv}, and \ref{tab:targetsK2nonosc}  share the following elements. Additional content is explained in the table notes of each table. 
The first vertical panel of the tables provides various stellar identifiers.

\begin{itemize}
    \item The star sequence number in the SB9 catalogue and the identifier in the {TESS} Input Catalogue (TIC) are provided in the first two columns.
    \item Additionally, in Table\,\ref{tab:targetsKepler} and \ref{tab:targetsK2nonosc}, the TIC identifier is followed by the star identifyer in the \textit{Kepler} Input Catalogue (KIC), {Ecliptic Input Catalogue (EPIC)}, respectively. 
    Table\,\ref{tab:BriteRv} provides an alternative identifier for the bright stars in the Bayer or Flamesteed catalogue.
\end{itemize}

The next vertical panel provides information on the observations. Further literature values are provided for a better characterization of the objects.
\begin{itemize}
    \item The next column provides information on the length of the data set. \Table{tab:targetsTESS} lists the number of {TESS Sectors} with the typical length of 27.5\,d each, or 
    \Table{tab:targetsKepler} provides the number of the so called \textit{Kepler} Quarters with the typical length of 90\,d, each.
    As \Table{tab:BriteRv} summarizes various results, we refer the reader to the notes below the table for more details. 
    \item The apparent magnitude in Johnson V, as reported in the Simbad Catalog, is given.
    \item The column T$_\mathrm{eff}$ gives the effective temperature of the star in Kelvin.
\end{itemize}  

The third vertical panel reports the observed and derived seismic quantities and the period of long periodic variations
\begin{itemize}
    \item The column labled as '\num' and '\dnu' report the peak frequency of the oscillation-power excess and large-frequency separation with their respective uncertainties. Because the giant is by far the brighter component, we can safely assume that the oscillations originate from the primary. In case no oscillation pattern could be detected, the field is filled with a '$-$'. The seismic extraction pipeline A2Z is underestimating the uncertainties for luminous red-giants. In a conservative approach, we decided only to report the value but not provide an uncertainty for them.
    \item The columns $M$, $R$, and $\log$$g$ report the seismically inferred values for the mass, radius and surface gravity of the primary star in solar units.  ($\nu_{\mathrm{max},\odot}$=3100\,$\mu$Hz, $\Delta\nu_\odot$=135.2\,$\mu$Hz and 5777\,K for the effective temperature of the Sun).  

\end{itemize}

The last five columns report the orbital parameters as given in the SB9 catalogue.
\begin{itemize}
    \item P$_\mathrm{orb}$ and $e$ describe the period and eccentricity of the orbit, respectively. 
    \item K$_1$ reports the radial velocity amplitude of the primary stellar component. K$_2$ reports the RV amplitude of the secondary, if known. No uncertainties are provided in the SB9 for K$_1$~and~K$_2$. 
    \item The last column reports the Grade of the solution, reported by SB9. It ranges between poor (grade = 0) and definitive (5). If no value was available, the field is filled by $-$.
\end{itemize}

\begin{sidewaystable*}%[t!]
\caption{Catalogue and parameters for oscillating red-giant binaries characterized through photometry from the \Kepler and K2 mission. \label{tab:targetsKepler}}
    \centering
\tabcolsep=1pt
    \begin{tabular}{rrrr|rrr|rrrrr|rrrrr}
    \hline\hline
%%%%%%%%%%%%%%%%%%%%%%%%%%%%%%%%%%%%%%%%%%%%%%%%%%%%%%%%%%%%%%%%%%%%%%%%%%%%%%%%%%%%%%%%%%%%%%%%
\multicolumn{1}{c}{SB9} &       \multicolumn{1}{c}{TIC} &       \multicolumn{1}{c}{KIC} &       &       \multicolumn{1}{c}{Q}   &       \multicolumn{1}{c}{V}   &       \multicolumn{1}{c}{T$_\mathrm{eff}$}                            &       \multicolumn{1}{c}{$\nu\mathrm{max}$}                   &       \multicolumn{1}{c}{$\Delta\nu$}                 &                       \multicolumn{1}{c}{M}                   &       \multicolumn{1}{c}{R}                   &       \multicolumn{1}{c}{$\log$$g$}                   &       \multicolumn{1}{c}{P$_\mathrm{orbit}$}                  &       \multicolumn{1}{c}{$e$}                 &       \multicolumn{1}{c}{K$_1$}       &       \multicolumn{1}{c}{K$_2$}       &       \multicolumn{1}{c}{SB9} \\
\multicolumn{1}{c}{Seq} &               &               &               &               &       \multicolumn{1}{c}{[mag]}       &       \multicolumn{1}{c}{[K]}                         &       \multicolumn{1}{c}{[$\mu$Hz]}                   &       \multicolumn{1}{c}{[$\mu$Hz]}                   &                       \multicolumn{1}{c}{[M/M$_\odot$]}                       &       \multicolumn{1}{c}{[R/R$_\odot$]}                       &       \multicolumn{1}{c}{[dex]}                       &       \multicolumn{1}{c}{[days]}                      &                       &       \multicolumn{1}{c}{[km/s]}      &       \multicolumn{1}{c}{[km/s]}      &       \multicolumn{1}{c}{Grd} \\[2mm] \hline

3242    &       139109614       &       KIC\,5023931    &       *.**    &       15      &       13.32   &       4728    $\pm$   92      (A)     &       51.9    $\pm$   1.1     &       4.88    $\pm$   0.10    &                        1.7     $\pm$   0.1     &       10.6    $\pm$   0.4     &       2.62    $\pm$   0.01    &       209.89  $\pm$   0.04    &       0.585   $\pm$   0.007   &       20.5    &       $-$     &       $-$     \\
3241    &       139154243       &       KIC\,5112840    &               &       12      &       13.84   &       5053    $\pm$   90      (A)     &       112.1   $\pm$   1.9     &       8.85    $\pm$   0.35    &                        1.8     $\pm$   0.2     &       7.2     $\pm$   0.4     &       2.97    $\pm$   0.01    &       5200    $\pm$   120     &       0.1     $\pm$   0.04    &       3.44    &       $-$     &       $-$     \\
3243    &       139109487       &       KIC\,5024240    &       *       &       14      &       14.32   &       4990    $\pm$   108     (A)     &       145.8   $\pm$   2.2     &       12.05   $\pm$   0.26    &                        1.1     $\pm$   0.1     &       5.1     $\pm$   0.2     &       3.08    $\pm$   0.01    &       66.837  $\pm$   0.016   &       0.22    $\pm$   0.03    &       4.04    &       $-$     &       $-$     \\
3233    &       184011023       &       KIC\,5024851    &       *       &       15      &       11.69   &       4124    $\pm$   68      (A)     &       3.9     $\pm$   0.7     &       0.65    $\pm$   0.10    &                        1.9     $\pm$   0.8     &       41.8    $\pm$   11.7    &       1.47    $\pm$   0.07    &       2379    $\pm$   9       &       0.24    $\pm$   0.04    &       5.76    &       $-$     &       $-$     \\
3234    &       139109185       &       KIC\,4937056    &       *       &       14      &       13.12   &       4848    $\pm$   85      (A)     &       48.3    $\pm$   3.8     &       5.06    $\pm$   0.09    &                        1.3     $\pm$   0.2     &       9.4     $\pm$   0.8     &       2.59    $\pm$   0.03    &       2920    $\pm$   30      &       0.39    $\pm$   0.03    &       6.01    &       $-$     &       $-$     \\
3236    &       139109401       &       KIC\,5023953    &       *       &       15      &       12.94   &       4872    $\pm$   86      (A)     &       49.9    $\pm$   2.0     &       4.68    $\pm$   0.04    &                        1.9     $\pm$   0.1     &       11.2    $\pm$   0.5     &       2.61    $\pm$   0.02    &       7369    $\pm$   19      &       0.608   $\pm$   0.012   &       4.28    &       $-$     &       $-$     \\
3237    &       139154396       &       KIC\,5112361    &       *       &       15      &       13.31   &       4813    $\pm$   90      (A)     &       66.1    $\pm$   4.0     &       6.16    $\pm$   0.17    &                        1.4     $\pm$   0.2     &       8.6     $\pm$   0.6     &       2.73    $\pm$   0.03    &       1449    $\pm$   4       &       0.06    $\pm$   0.03    &       5.13    &       $-$     &       $-$     \\
3231    &       139109502       &       KIC\,5024476    &               &       15      &       12.83   &       5054    $\pm$   91      (A)     &       66.4    $\pm$   1.6     &       5.66    $\pm$   0.13    &                        2.1     $\pm$   0.1     &       10.4    $\pm$   0.4     &       2.74    $\pm$   0.01    &       1524    $\pm$   5       &       0.24    $\pm$   0.04    &       5.37    &       $-$     &       $-$     \\
3239    &       139153844       &       KIC\,5024414    &       *       &       14      &       12.95   &       5042    $\pm$   99      (A)     &       78.5    $\pm$   3.9     &       6.36    $\pm$   0.17    &                        2.2     $\pm$   0.2     &       9.7     $\pm$   0.6     &       2.81    $\pm$   0.02    &       3360    $\pm$   50      &       0.72    $\pm$   0.06    &       4.7     &       $-$     &       $-$     \\
3247    &       184011491       &       KIC\,4937775    &       *       &       8       &       13.45   &       5093    $\pm$   164              &       89.2    $\pm$   7.7     &       7.36    $\pm$   0.15    &                        1.9     $\pm$   0.3     &       8.3     $\pm$   0.8     &       2.87    $\pm$   0.04    &       1240    $\pm$   30      &       0.35    $\pm$   0.06    &       4       &       $-$     &       $-$     \\
3246    &       139153963       &       KIC\,5024582    &       *       &       14      &       13.01   &       4803                     (A)     &       47.2    $\pm$   4.3     &       5.00    $\pm$   0.29    &                        1.2     $\pm$   0.2     &       9.3     $\pm$   1.1     &       2.58    $\pm$   0.04    &       1584    $\pm$   10      &       0.39    $\pm$   0.06    &       3.94    &       $-$     &       $-$     \\
3101    &       405717854       &       KIC\,11753949   &               &       14      &       6.43    &       4336    $\pm$   71      (A)     &       11.3    $\pm$   1.5     &       1.65    $\pm$   0.19    &                        1.2     $\pm$   0.4     &       19.4    $\pm$   4.1     &       1.94    $\pm$   0.05    &       674.7   $\pm$   0.5     &       0.022   $\pm$   0.004   &       7.81    &       $-$     &       $-$     \\
3515    &       27186427        &       KIC\,11408263   &               &       18      &       6.46    &       4840    $\pm$   74               &       41.8    $\pm$   4.5     &       4.52    $\pm$   0.57    &                        1.3     $\pm$   0.4     &       10.2    $\pm$   2.1     &       2.53    $\pm$   0.04    &       4204    $\pm$   10      &       0.3898  $\pm$   0.0086  &       4.7     &       $-$     &       $-$     \\
2660    &       63289148        &       KIC\,9528112    &               &       17      &       7.18    &       3313    $\pm$   144              &       28.8    $\pm$   2.9     &       3.43    $\pm$   0.08    &                        0.7     $\pm$   0.1     &       10.1    $\pm$   1.1     &       2.29    $\pm$   0.04    &       926.3   $\pm$   36.4    &       0.08    $\pm$   0.2     &       1.8     &       $-$     &       $-$     \\
3235    &       139153978       &       KIC\,5112741    &       g       &       17      &       12.76   &       4877    $\pm$   72               &                               &                               &                                                &                               &                                &       17.6978 $\pm$   0.0003  &       0.022   $\pm$   0.012   &       42.7    &       $-$     &       $-$     \\
1151    &       350739215       &       KIC\,11913210   &       g       &       3       &       7.08    &       3709    $\pm$   366              &                               &                               &                                                &                               &                                &       5750                    &       0.29                     &       6.8     &       $-$     &       2       \\ \hline

1827    &       437039582       &       EPIC\,211394348 &               &       1       &       12.03   &       4930    $\pm$   82               &       119.0   $\pm$   6.0     &       9.90    $\pm$   0.95    &                        1.3     $\pm$   0.3     &       6.1     $\pm$   0.9     &       2.99    $\pm$   0.02    &       1233    $\pm$   19      &       0.13    $\pm$   0.05    &       4.4     &       $-$     &       $-$     \\
1925    &       175290697       &       EPIC\,211976270 &               &       2       &       6.92    &       5040    $\pm$   54               &       76.0    $\pm$   7.0     &       6.57    $\pm$   0.34    &                        1.8     $\pm$   0.3     &       8.8     $\pm$   1.0     &       2.80    $\pm$   0.04    &       994.4   $\pm$   1.2     &       0.806   $\pm$   0.004   &       9.8     &       $-$     &       $-$     \\
2586    &       14602163        &       EPIC\,211993818 &               &       2       &       7.38    &       5716    $\pm$   72               &       60.0    $\pm$   7.0     &       5.81    $\pm$   2.87    &                        1.8     $\pm$   1.8     &       9.6     $\pm$   6.8     &       2.72    $\pm$   0.05    &       3900    $\pm$   17      &       0.61    $\pm$   0.007   &       18.75   &       $-$     &       $-$     \\ \hline
\end{tabular}
\tablefoot{See Appendix\,\ref{appendix:tables} for a detailed description of each column of the table. $Q$ gives the numbers of Quarters of \Kepler data were obtained. Two flags are used to indicate special characteristics of the systems, described in the paper. Systems which are confirmed members of NGC\,6819 are indicated with a the flag '*' besides the KIC identifier. 
The label 'g' stands for non-oscillating stars, which were confirmed to be giants from their clear granulation signature in the PSD of \Kepler data. 
The star that falls into the described red-giant desert are marked through the flag '**'. Temperatures which are taken from the Apogee catalogue are flagged with an (A).}

\vspace{15mm}

\caption{Literature values for system seismically characterized from BRITE photometry or radial-velocity studies.}
    \centering
\tabcolsep=1pt
    \begin{tabular}{rrr|rrr|rrrrrrrrrr}
\hline\hline
\multicolumn{1}{c}{SB9} &       \multicolumn{1}{c}{HD}  &       \multicolumn{1}{c}{Alt.ID}                      &       \multicolumn{1}{c}{Instrument}  &       \multicolumn{1}{c}{V}   &       \multicolumn{1}{c}{T$_\mathrm{eff}$}    &       \multicolumn{1}{c}{Seismic}     &                       \multicolumn{1}{c}{P$_\mathrm{rot}$}    &       \multicolumn{1}{c}{M}   &       \multicolumn{1}{c}{R}   &       \multicolumn{1}{c}{$\log$$g$}   &       \multicolumn{1}{c}{P$_\mathrm{orbit}$}                  &       \multicolumn{1}{c}{$e$}                 &       \multicolumn{1}{c}{K$_1$}       &                       \multicolumn{1}{c}{Reference}   \\
\multicolumn{1}{c}{Seq} &               &                               &               &       \multicolumn{1}{c}{[mag]}       &       \multicolumn{1}{c}{[K]} &       \multicolumn{1}{c}{diagnostic}  &                       \multicolumn{1}{c}{[d]} &       \multicolumn{1}{c}{[M/M$_\odot$]}       &       \multicolumn{1}{c}{[R/R$_\odot$]}       &       \multicolumn{1}{c}{[dex]}       &       \multicolumn{1}{c}{[days]}                      &                               &       \multicolumn{1}{c}{[km/s]}      &                       \\[2mm] \hline
                                                                                                                                                                                                                                                                                                                        
2837    &       194317  &       39\,Cyg                 &       BRITE/UBr: 19\,d   &       4.44    &               &       $\nu_\mathrm{max}=$9.4$\pm$0.7\,$\mu$Hz &                       $-$     &       1.9\,$\pm$0.1   &       1.9\,$\pm$0.1   &       1.91\,$\pm$0.02 &       31292   $\pm$   324     &       0.495   $\pm$   0.023   &       3.23    &                       K19     \\
3489    &       101379  &       12\,Mus                 &       BRITE/BHr: 152\,d  &       5.1     &               &       $\tau_\mathrm{ACF}$= 5252\,min  &                       $-$     &       1.0\,$\pm$0.3   &       16.6\,$\pm$1    &       1.96\,$\pm$0.03 &       61.408  $\pm$   0.027   &       0.012   $\pm$   0.01    &       12.91   &                       K19     \\ \hline
                                                                                                                                                                                                                                                                                                                        
2601    &       28307   &       $\theta^1$\,Tau                 &       RV (MSC): 190\,d   &       3.84    &       5000 $\pm$ 150  &       $\nu_\mathrm{max}\simeq$90\,$\mu$Hz, $\Delta\nu$=6.9$\pm$0.2 &                       138.2   &       $\sim$2.7       &       $\sim$10        &               &       5939    $\pm$   46      &       0.57    $\pm$   0.022   &       7.17    &                       B15a    \\ \hline
    \end{tabular}
     \tablefoot{See Section\,\ref{appendix:tables} for more details on the columns of the table. In addition to the content described, the third column reports a more commonly known catalogue identifier. The last two columns report on the observational technique, and the analysis method. BRITE indicates photometric observations with the BRITE satellite. MSC-RV stands for multi-site campaign to measure the radial-velocity variations. While 39\,Cyg and $\theta^1$\,Tau exhibited significant oscillation mode amplitudes, which allowed the direct measurement of \num through a multicomponent fit, the seismic inferences were drawn from the granulation signal. K19: \cite{Kallinger2019}, B15a: \cite{Beck2015a}
     }
 \label{tab:BriteRv}

\end{sidewaystable*}

\begin{sidewaystable*}%[t!]

\caption{Catalogue and parameters for red-giant binaries characterized through photometry from the \Kepler and K2 mission for which no signature of oscillations were detected.}
    \centering
\tabcolsep=10pt
    \begin{tabular}{rrrr|rrr|rrrrrr|rrrrr}
                                                                                                        
\hline\hline                                                                                                                                                                    
%%%%%%%%%%%%%%%%%%%%%%%%%%%%%%%%%%%%%%%%%%%%%%%%%%%%%%%%%%%%%%%%%%%%%%%%%%%%%%%%%%%%%%%%%%%%%%%%
\multicolumn{1}{c}{SB9} &       \multicolumn{1}{c}{TIC} &       \multicolumn{1}{c}{KIC} &               &       \multicolumn{1}{c}{Obs} &       \multicolumn{1}{c}{V}   &       \multicolumn{1}{c}{T$_\mathrm{eff}$}    &       \multicolumn{1}{c}{P$_\mathrm{orbit}$}                  &       \multicolumn{1}{c}{$e$}                 &       \multicolumn{1}{c}{K$_1$}       &       \multicolumn{1}{c}{K$_2$}       &       \multicolumn{1}{c}{SB9} \\
\multicolumn{1}{c}{Seq} &               &               &               &               &       \multicolumn{1}{c}{[mag]}       &       \multicolumn{1}{c}{[K]} &       \multicolumn{1}{c}{[days]}                      &                               &       \multicolumn{1}{c}{[km/s]}      &       \multicolumn{1}{c}{[km/s]}      &       \multicolumn{1}{c}{Grade}       \\[2mm] \hline
3244    &       138970759       &       KIC\,5023822    &       **      &       14      &       14.97   &               &       40.744  $\pm$   0.008   &       0.586   $\pm$   0.009   &       14.21   &       $-$     &       $-$     \\
3248    &       139109639       &       KIC\,5024150    &               &       5       &       14.5    &               &       7810    $\pm$   60      &       0.67    $\pm$   0.04    &       3.8     &       $-$     &       $-$     \\
3277    &       139109632       &       KIC\,5024607    &               &       11      &       15.23   &       4750.0  &       414     $\pm$   3       &       0.71    $\pm$   0.08    &       13      &       $-$     &       $-$     \\
3264    &       184011237       &       KIC\,5024870    &               &       11      &       14.77   &               &       121.57  $\pm$   0.017   &       0.454   $\pm$   0.008   &       15.6    &       $-$     &       $-$     \\
3313    &       138970211       &       KIC\,5111815    &               &       14      &       15.11   &               &       3.5749                  &       0.01    $\pm$   0.005   &       76.3    &       97.3    &       $-$     \\
3310    &       139154299       &       KIC\,5112816    &               &       14      &       14.56   &               &       33.2393 $\pm$   0.0022  &       0.366   $\pm$   0.006   &       39.97   &       $-$     &       $-$     \\
3278    &       184010246       &       KIC\,5200656    &               &       14      &       15.2    &               &       370.3   $\pm$   2.1     &       0.11    $\pm$   0.04    &       12.8    &       $-$     &       $-$     \\
3258    &       184090973       &       KIC\,5201088    &               &       14      &       14.52   &       6000    &       135     $\pm$   0.3     &       0.28    $\pm$   0.03    &       10.3    &       $-$     &       $-$     \\
2660    &       63289148        &       KIC\,9528112    &               &       17      &       7.18    &               &       926.3   $\pm$   36.4    &       0.08    $\pm$   0.2     &       1.8     &       $-$     &       $-$     \\
                                                                                                                                                                                
525     &       307679747       &       EPIC\,212173112 &       osc? Ec &       2       &       10.8    &               &       10.173                  &       0                       &       32.2    &       $-$     &       1       \\
2586    &       14602163        &       EPIC\,211993818 &       osc?    &       2       &       7.38    &               &       3900    $\pm$   17      &       0.61    $\pm$   0.007   &       18.75   &       $-$     &       $-$     \\ 
                                                                                                                                                                                                                        
1813    &       437039105       &       EPIC\,211409376 &               &       3       &       12.57   &               &       10.0552 $\pm$   0.0001  &       0                       &       52.1    &       59.4    &       $-$     \\
1830    &       437034592       &       EPIC\,211385284 &               &       1       &       8.82    &       4750    &       1315    $\pm$   5       &       0.15    $\pm$   0.04    &       4.39    &       $-$     &       $-$     \\
1833    &       46305337        &       EPIC\,211427165 &       ell.var &       3       &       13.74   &       5500    &       2.8231  $\pm$   0       &       0                       &       60.6    &       86.2    &       $-$     \\
2957    &       188509972       &       EPIC\,206012818 &               &       1       &       7.78    &       3900    &       791.8   $\pm$   1.1     &       0.17    $\pm$   0.04    &       6.6     &       $-$     &       $-$     \\ \hline
%%%%%%%%%%%%%%%%%%%%%%%%%%%%%%%%%%%%%%%%%%%%%%%%%%%%%%%%%%%%%%%%%%%%%%%%%%%%%%%%%%%%%%%%%%%%%%%%

    \end{tabular}
    \tablefoot{See Appendix\,\ref{appendix:tables} for a detailed description of each column of the table. }
 \label{tab:targetsK2nonosc}

\end{sidewaystable*}

\end{appendix}
\end{document}